\documentclass{article}

\usepackage{arxiv}
\usepackage[utf8]{inputenc} 
\usepackage[T1]{fontenc}    
\usepackage{hyperref}       
\usepackage{url}            
\usepackage{booktabs}       
\usepackage{amsfonts}       
\usepackage{nicefrac}       
\usepackage{microtype}      
\usepackage{lipsum}
\usepackage{graphicx}

\usepackage[english]{babel}
\usepackage{amsmath}
\usepackage{amssymb}
\usepackage{amsthm}
\usepackage{pdfpages}
\usepackage{float} 
\usepackage{tikz}
\usetikzlibrary{patterns, decorations}
\usetikzlibrary{decorations.pathreplacing, calc}
\usepackage{caption}
\usepackage{subcaption}
\usepackage{xcolor}
\usepackage[linesnumbered,ruled]{algorithm2e} 
\usepackage{natbib}
\usepackage[bibliography=common]{apxproof}
\usepackage{nicematrix}
\usepackage{layouts} 
\usepackage{xcolor}
\hypersetup{
    colorlinks,
    linkcolor={black!50!black},
    citecolor={blue!50!black},
    urlcolor={blue!80!black}
}

\newtheoremrep{theorem}{Theorem}

\newcommand\norm[1]{\left\lVert#1\right\rVert}
\newcommand{\proofinappendixstatement}{
    \begin{proof}
        The proof is given in the appendix.
    \end{proof}
    }

\title{Spatially smoothed robust covariance estimation for local outlier detection}
\author{
  Patricia Puchhammer \\
  TU Wien\\
  \texttt{patricia.puchhammer@tuwien.ac.at} \\
   \And
 Peter Filzmoser \\
  TU Wien\\
  \texttt{peter.filzmoser@tuwien.ac.at} \\
}

\begin{document}

\maketitle
\begin{abstract}
Most multivariate outlier detection procedures ignore the spatial dependency of observations, which is present in many real data sets from various application areas. This paper introduces a new outlier detection method that accounts for a (continuously) varying covariance structure, depending on the spatial neighborhood of the observations. The underlying estimator thus constitutes a compromise between a unified global covariance estimation, and local covariances estimated for individual neighborhoods. Theoretical properties of the estimator are presented, in particular related to robustness properties, and an efficient algorithm for its computation is introduced. The performance of the method is evaluated and compared based on simulated data and for a data set recorded from Austrian weather stations.
\end{abstract}
\keywords{Spatial Outlier Detection \and MRCD \and Pairwise Mahalanobis Distance}

\section{Introduction}
\label{sec:intro}

The identification of multivariate outliers is probably one of the most important tasks in multivariate data analysis. 
A need to find outliers in order to make further analyses more reliable,
or the direct interest in the outliers themselves motivate the numerous approaches available for multivariate outlier detection.
The identified outliers are supposed to deviate to a certain extent 
from the main trend or structure of the data majority, and thus they are
also called ``global outliers'' \citep{Filzmoser2013}.
In contrast, the term ``local outliers'' refers to a setting where additional
information regarding some kind of neighborhood is available, for example provided by
spatial coordinates of the observations.
Then, local outliers are observations which clearly differ from the 
multivariate measurements of their spatial neighbors indicating local anomalies that spark interest and make further analysis essential.
Nevertheless, the values themselves might still be in an ordinary range of the data set, and thus the observation would not be outlying in a global sense. 

Existing statistical approaches for multivariate local outlier detection are often based on a distance measure and neighborhood structure. 
A neighborhood $a$ is defined as a subset of the set of observation indexes,
say $\{1, \ldots , n\}$. 
A $p$-variate observation $\boldsymbol{x}_i$, for $i\in \{1, \ldots , n\}$,
is defined to be in neighborhood $a$ if and only if $i \in a$. 
The decision if some observation $\boldsymbol{x}$ is in a neighborhood is typically based on its spatial coordinates $s(\boldsymbol{x})$. One way to construct the spatial neighborhood is to take a spatial k-nearest-neighborhood of each point $\boldsymbol{x}$, where $k \in \mathbb{N}$. For a fixed $\boldsymbol{x}$, the spatial distance to another point $\boldsymbol{y}$ is defined as the Euclidean distance 
\begin{align*}
    d_{\boldsymbol{x}}(\boldsymbol{y}) = \norm{s(\boldsymbol{x})-s(\boldsymbol{y})} = \left[\left(s(\boldsymbol{x})-s(\boldsymbol{y})\right)' \left(s(\boldsymbol{x})-s(\boldsymbol{y})\right)\right]^{1/2}.
\end{align*}
A spatial neighborhood $a_k(\boldsymbol{x})$ can then be defined as the set of the $k$ many spatial nearest observations, 
\begin{align}
\label{eq:spatially_knn_def}
    a_k(\boldsymbol{x}) = \{\boldsymbol{x}_j: \ d_{\boldsymbol{x}}(\boldsymbol{x}_j) \le  d_{\boldsymbol{x}(k)}\},
\end{align}
where $d_{\boldsymbol{x}(1)} \le d_{\boldsymbol{x}(2)} \le \ldots \le d_{\boldsymbol{x}(k)} \le d_{\boldsymbol{x}(n)}$ are the sorted distances to all observations $\boldsymbol{x}_i, i = 1, \ldots, n$.
Regarding local outlier detection, a distance measure often used to evaluate outlyingness is the pairwise Mahalanobis distance~(MD).   
For a neighborhood $a$ with covariance $\boldsymbol{\Sigma}_a$ and an observation $\boldsymbol{x}_i$ in neighborhood $a$, the pairwise MD is defined as
\begin{align*}
    \text{MD}_{\boldsymbol{\Sigma}_a} (\boldsymbol{x}_i, \boldsymbol{x}_j) = \left[ (\boldsymbol{x}_i - \boldsymbol{x}_j)' \boldsymbol{\Sigma}_a ^{-1} (\boldsymbol{x}_i - \boldsymbol{x}_j) \right] ^{1/2} \ \ \ \ \text{for all } j \in a.
\end{align*} 
In general, the MD describes the distance between two observations, where the Euclidean distance in the feature space is adapted according to local distribution properties.

The estimation method used to determine $\boldsymbol{\Sigma}_a$ for all neighborhoods is key to good and reliable results. 
It is essential that outlying observations themselves are not affecting the estimation, since this could possibly mask outliers, leaving them undetected.
Thus, a robust covariance estimation on the neighborhood level is necessary.
One of the most widely used robust estimators for the covariance is the Minimum Covariance Determinant (MCD) estimator \citep{rousseeuw1984least, Rousseeuw1985}, where one has to identify the $h$ sub-sample of observations 
(where $h$ is fixed e.g.~to half of the observations) that minimize the determinant of its sample covariance. The MCD covariance estimator is then
given by the sample covariance of the $h$ subset, multiplied by a 
consistency factor \citep{Croux1999}. For its computation, 
the fastMCD algorithm developed by \cite{Rousseeuw1999} introduces an iterative concentration step, so-called C-step, that guarantees a decrease of the objective function until convergence to a (local) minimum, making the MCD estimator faster and even more popular.
The global minimum is then approximated by iterating for a number of random starting values and choosing the smallest local minimum.
By selecting a small number of good deterministic starting values for the fastMCD, the detMCD algorithm from \cite{Hubert2012} improves run-time even more.
In spite of these excellent features of the MCD estimator,
as well as affine equivariance and high robustness,
one drawback is that the concept is not applicable in case of singularity
of the sample covariance matrix of the $h$ subset, which can easily occur. As for many methods, regularity of the estimated covariance is also needed to compute Mahalanobis distances.
Especially in a setting where we are restricted to local neighborhoods consisting of a possibly small subset of observations, we might have a situation where regularity cannot be achieved and thus an inversion of the local covariance matrix is not possible.
One solution is to base the local estimation on regularized robust covariance estimators, such as the recently developed Minimum Regularized Covariance Determinant (MRCD) estimator from \cite{Boudt2020} (or also on the Fritsch estimator, \cite{Fritsch2012}).
One of the many attractive properties of the MRCD is that a slightly adapted fastMCD algorithm based on C-steps is also applicable.

Existing methods for local outlier detection have different ways to 
define the covariance matrix in order to ensure regularity. 
The method of \cite{Filzmoser2013} is dealing with regularity issues by using the MCD estimator calculated on the whole data set, i.e.,~$\boldsymbol{\Sigma}_a = \boldsymbol{\Sigma}$, thus imposing a global covariance structure. 
They have shown that for i.i.d.~Gaussian random vectors 
$\boldsymbol{x}_1,\ldots ,\boldsymbol{x}_n$ with mean
$\boldsymbol{\mu}$ and covariance matrix
$\boldsymbol{\Sigma}$, the conditional distribution of 
the pairwise squared MD$_{\boldsymbol{\Sigma}}(\boldsymbol{x}_i,\boldsymbol{x}_j)$, for $j=1,\ldots ,n$, given $\boldsymbol{x}_i$, is a
non-central chi-square distribution with $p$ degrees of freedom and
non-centrality parameter MD$^2_{\boldsymbol{\Sigma}}(\boldsymbol{x}_i,
\boldsymbol{\mu})$.
Instead of a fixed cut-off value for the pairwise MD, different sophisticated visual approaches are used. 
By plotting a degree of isolation based on the pairwise MD and quantiles of the non-central chi-square distribution, suspicious and highly isolated observations can be discovered and analysed in more detail.
Quite contrary to \cite{Filzmoser2013}, the method of \cite{Ernst2016} uses a very local covariance estimation by taking individual k-nearest-neighbor (kNN) neighborhoods for each point separately into account.
To tackle the regularity issues, they use a regularized covariance estimation (originally the estimator from \cite{Fritsch2012}, for the MRCD see also the adaptation made in \cite{Bellino2019}) for each individual kNN neighborhood.
Additionally, they introduce the concept of the ``next distance'', which is also MD based, and use the upper fence of the adjusted 
boxplot of \cite{Hubert2008} of all next distances as a non-parametric cut-off value for detecting outliers. 

Since it is not necessary to use a MD concept to find local outliers, a short detour to machine learning techniques might be interesting.
One of the most prominent approaches for detecting multivariate local outliers in machine learning is the local outlier factor (LOF) developed by \cite{Breunig2000}. 
Initially, locality refers to multivariate values and Euclidean distances in the feature space but this method can also be canonically adapted to spatial local outlier detection. 
In \cite{Schubert2012} this adaptation and further LOF-based approaches are discussed.
Interestingly, also LOF and its variants are based on a concept of distance and neighborhoods.

The existing methods have shortcomings in various ways that have not yet been properly addressed. 
The rather global nature of the method of \cite{Filzmoser2013} leads to a reliable and robust estimation of the covariance. 
Nevertheless, it is somewhat questionable if the estimated covariance is applicable and representative for the covariance structure on the local level. 
Trying to solve this issue of missing locality in the estimation, \cite{Ernst2016} resort to a very local approach by recalculating the covariance matrix for each observation separately. 
Although more locality is achieved, the method is not taking into account that covariance matrices are not likely to change abruptly from one neighborhood
to a next one.
Also, the number of estimated parameters is extremely high and based on rather few observations, even if the local covariance structure is abruptly changing.
A more global estimation of the local covariance matrices might be more stable and reliable and might also avoid repetitive calculations. 
It seems that until now there are only two extremes regarding locality of the covariance estimation available.

We bridge the gap between the fully global and the fully local approach by providing a covariance estimator based on the MRCD that addresses the missing locality on the one hand and the missing spatial smoothness on the other.
By providing the possibility to set the amount of spatial smoothing and the size of the neighborhoods we get a generalization of the two detection methods, with the goal that good outlier detection properties based on the new local covariance estimations are achieved.
Moreover, the covariance estimate can be seen as a generalization of the MRCD when the data set has additional sub-structures.

The paper is organized as follows. 
In Section 2 we introduce the new covariance estimator, derive its properties as well as properties of the original MRCD and establish the methodology to detect local outliers. 
An algorithm and the derivation of a generalized C-step are discussed in Section 3.
Section 4 provides simulation results regarding run time, convergence and outlier detection, while in section 5, a real data set is analyzed using the newly developed local outlier detection method.
Finally, the main results are presented and summarized in the conclusions.

\section{Methodology}
\label{sec:method}

\subsection{Spatially smoothed MRCD estimators}
Assume that the $p$-dimensional observations 
$\boldsymbol{x}_i = (x_{i1}, \ldots, x_{ip})'$, for $i=1,\ldots ,n$,
are arranged as rows in the $n \times p$ matrix $\boldsymbol{X}$ with $n > 2p$. 
Furthermore, each observation has spatial information available, e.g. spatial coordinates, and is assigned to one of $N$ many neighborhoods, defined by
the index sets
$a_1, \ldots , a_N$ of size $n_1, \ldots, n_N$.

The goal of the proposed method is to obtain local covariance estimates for each neighborhood that are suitable for calculating the pairwise MDs
and to some extent smooth among nearby neighborhoods.
Since the MCD estimator requires at least $n > 2p$ to provide a regular covariance matrix, which is a severe limitation especially for 
small neighborhoods, we focus on its regularized extension, the MRCD estimator.
Instead of minimizing the determinant of the sample covariance matrix as in the MCD case, the minimization objective is the determinant of a convex combination of the sample covariance and a symmetric and positive definite matrix, the so-called target matrix $\boldsymbol{T}$. 
\cite{Boudt2020} suggest a data-driven approach based on the condition number of the covariance matrix to set the degree of regularization $\rho$ which is used in the convex combination.
Regarding the target matrix $\boldsymbol{T}$, it is sensible to choose a robust and regular covariance, e.g.~a diagonal matrix based on univariate robust scale estimates.

In the following, we adapt the idea of the MRCD estimator to our setting of local and smooth covariance estimation.
Let $\mathcal{H} = (H_1, \ldots , H_N) $ be subsets of the index sets 
$a_1, \ldots , a_N$ defining the neighborhoods. 
The size of each subset $H_i$
is $h_i = |H_i|= \lceil \alpha n_i \rceil$, for $i = 1, \ldots , N$,
where $\alpha$ is selected in the interval $[0.5,1]$, and 
$\lceil \cdot \rceil$ is the ceiling function, rounding up to the next integer. 
A smaller value of $\alpha$ will result in more robustness against outliers,
and it would also be possible to adjust this value to each neighborhood individually. 
The observations of subset $H_i$ are written as matrix $\boldsymbol{X}_{H_i}$,
with dimensionality $h_i\times p$.
Let the neighborhood specific MRCD-based covariance matrix $\boldsymbol{K}_i(\mathcal{H})$, for $i = 1, \ldots , N$, be defined as
\begin{align}
    \label{eq:K_def}
    \boldsymbol{K}_i(\mathcal{H})= \rho_i \boldsymbol{T} + (1-\rho_i) c_{\alpha} Cov(\boldsymbol{X}_{H_i}),
\end{align}
with $Cov(\boldsymbol{Y})$ being the sample covariance matrix of $\boldsymbol{Y}$, and $c_{\alpha}$ a consistency factor for the proportion $\alpha$ \citep[see][]{Croux1999}.
The regularization parameter $\rho_i$ is set individually for each neighborhood,
and it could also be chosen as zero if the estimated covariance matrix 
is already invertible.
Finally, since we want to smooth the covariance matrices, it seems 
counter intuitive to choose neighborhood specific target matrices, 
which would also require more parameter estimations.
Therefore, we assume a global target matrix $\boldsymbol{T}$, 
taken as a robust and regular covariance matrix estimated based 
on the full data set $\boldsymbol{X}$. 
Since we assume $n>2p$ we propose to use the MCD estimator for $\boldsymbol{X}$ as target matrix.

We want to find the combination of subsets in $\mathcal{H}$ that minimizes the objective function
\begin{equation}
	\label{eq:objective_function}
	f(\mathcal{H}) = \sum_{i = 1}^{N}  \det\left( (1-\lambda) \boldsymbol{K}_i(\mathcal{H}) + \lambda \sum_{j = 1 , j \neq i}^{N} \omega_{ij} \boldsymbol{K}_j(\mathcal{H}) \right).
\end{equation}
\noindent
The tuning parameter $\lambda \in [0,1]$ is used to balance the influence of 
an individual local neighborhood and the remaining neighborhoods in the covariance estimations. 
In case of $\lambda = 0$, there is no spatial influence at all which is equivalent to the estimation of the MRCD for each neighborhood separately while using a global target matrix.  
For the other extreme $\lambda = 1$, the covariance matrix in a specific neighborhood is an average over the surrounding covariance estimates without adding local information from the neighborhood itself. Moreover, it is possible to interpret the second part in the determinant as a penalization term. Due to the minimization of the determinant, observations from $a_i$ that match well with the main trend of observations in neighborhoods with positive weights $\omega_{ij}$ are more likely to be in the optimal H-set if $\lambda$ is increased. 
The weights $\omega_{ij}$ are supposed to be non-negative, and
we set $\omega_{ii}=0$.
The elements of the weight vector
$\boldsymbol{\omega}_i=(\omega_{i1},\ldots ,\omega_{iN})'$
indicate the relative influence that the 
estimated covariances of other neighborhoods have on the covariance estimation
of the $i$-th neighborhood. 
Also, each weight vector has to sum up to one, $\sum_{j = 1}^{N} \omega_{ij} = 1$ for all $i = 1, \ldots, N$. 
All these weights need to be pre-specified, for example based on inverse geographical distances, and
are collected as rows in the weighting matrix $\boldsymbol{W} \in \mathbb{R}^{N \times N}$. 

Note that for the objective function (\ref{eq:objective_function}) a global 
minimum $\mathcal{H}^* = (H^*_i)_{i=1,\ldots ,N}$ exists since its domain consists of a finite number of subset combinations.
For this global minimum, the estimated covariance matrix for each neighborhood
$a_i$ is 
\begin{align}
    \hat{\boldsymbol{\Sigma}}_{SSM, i} = (1-\lambda) \boldsymbol{K}_i(\mathcal{H}^*) + \lambda \sum_{j = 1 , j \neq i}^{N} \omega_{ij} \boldsymbol{K}_j(\mathcal{H}^*),
\end{align}
and the location estimate $\hat{\boldsymbol{\mu}}_{SSM, i}$ is the sample mean of the selected observations $\boldsymbol{X}_{H^*_i}$.
We call these estimators the spatially smoothed MRCD (ssMRCD) location
and covariance estimators.

Although the neighborhood structure and the value of $\lambda$ will often depend on the data at hand, there are some sensible and natural choices for $\boldsymbol{W}$. If we have a neighborhood structure that has no further meaning and might just be used to divide the spatial space into subsets, an inverse-distance based weight matrix, also used in Sections~\ref{sec:Numerical_simulations} and \ref{sec:examples}, might be a good choice. Other possibilities include binary matrices with ones if neighborhoods share a border and zero otherwise, with rows scaled appropriately. Moreover, the regularity parameters can be set by default.
For a neighborhood $a_i$ we suggest to set the regularization parameter $\rho_{i}$ as the data driven value that is proposed by the MRCD algorithm in \cite{Boudt2020}, when interpreting the neighborhood as its own data set.

\medskip
\noindent
\subsection{Theoretical properties}

In the following we will show that the spatially smoothed MRCD estimators proposed here
are -- in contrast to the original MRCD estimator --
affine equivariant, and we derive its breakdown point. 
For a  data set $\boldsymbol{X} \in \mathbb{R}^{n \times p}$, 
a location and covariance estimator
are called affine equivariant for all neighborhoods if for any non-singular 
matrix $\boldsymbol{A} \in \mathbb{R}^{p \times p}$, any vector $\boldsymbol{b} \in \mathbb{R}^{p}$, and all $i = 1, 2, \ldots ,N$, it holds that
\begin{align}
	\label{eq:def_affineequivariant} 
	\hat{\boldsymbol{\mu}}_{SSM, i}(\boldsymbol{XA}' + \boldsymbol{1}_n \boldsymbol{b}') &= \hat{\boldsymbol{\mu}}_{SSM, i}(\boldsymbol{X})\boldsymbol{A}' + \boldsymbol{b}', \\
	\hat{\boldsymbol{\Sigma}}_{SSM, i}(\boldsymbol{XA}' + \boldsymbol{1}_n \boldsymbol{b}') &= \boldsymbol{A\hat{\Sigma}}_{SSM, i}(\boldsymbol{X})\boldsymbol{A}' . \nonumber
\end{align}

\medskip
\noindent

\begin{theoremrep}[Affine equivariance]
\label{AffineEqui_theorem}
Let $\boldsymbol{T}$ be any robust, regular and affine equivariant estimate of the covariance for the data set $\boldsymbol{X}$,
here denoted as $\boldsymbol{T}(\boldsymbol{X})$.
Then, the spatially smoothed MRCD estimators of location and covariance
with target matrix $\boldsymbol{T}(\boldsymbol{X})$ are affine equivariant. 
\end{theoremrep}

\begin{proof}
Let $i$ be fixed, $\boldsymbol{T}(\boldsymbol{X})$ be 
an estimator of covariance as described above and $\boldsymbol{Y} := \boldsymbol{XA}' + \boldsymbol{1}_n \boldsymbol{b}'$ 
be the transformed data matrix. Then, for any subset combination $\mathcal{H}$ and for all $j = 1, \ldots , N$ it holds that
\begin{align*}
	\boldsymbol{K}_j^Y (\mathcal{H}) &= \rho_j \boldsymbol{T}(\boldsymbol{Y}) + (1-\rho_j) Cov(\boldsymbol{Y}_{H_j})  \nonumber \\ 
	&= \rho_j \boldsymbol{T}(\boldsymbol{XA}' + \boldsymbol{1}_n\boldsymbol{b}) + (1-\rho_j) Cov(\boldsymbol{X}_{H_j}\boldsymbol{A}' + \boldsymbol{1}_n\boldsymbol{b})  \nonumber \\ 
	&= \rho_j \boldsymbol{A}\boldsymbol{T}(\boldsymbol{X})\boldsymbol{A}' + (1-\rho_j) \boldsymbol{A}Cov(\boldsymbol{X}_{H_j})\boldsymbol{A}' \nonumber \\
	&= \boldsymbol{A} \left[\rho_j \boldsymbol{T}(\boldsymbol{X}) + (1-\rho_j) Cov(\boldsymbol{X}_{H_j}) \right]\boldsymbol{A}' \nonumber \\
	&= \boldsymbol{A K}_j^X(\mathcal{H}) \boldsymbol{A}'.
\end{align*}
It follows that
\begin{align}
	\label{eq:proof_affineequi}
	&\left( (1-\lambda) \boldsymbol{K}_i^Y(\mathcal{H}) + \lambda \sum_{j = 1 , j \neq i}^{N} \omega_{ij} \boldsymbol{K}_j^Y(\mathcal{H}) \right) = \nonumber \\
	&= \left( (1-\lambda) \boldsymbol{AK}_i^X(\mathcal{H})\boldsymbol{A}' + \lambda \sum_{j = 1 , j \neq i}^{N} \omega_{ij} \boldsymbol{AK}_j^X(\mathcal{H})\boldsymbol{A}' \right) \nonumber \\
	&= \boldsymbol{A} \left( (1-\lambda) \boldsymbol{K}_i^X(\mathcal{H}) + \lambda \sum_{j = 1 , j \neq i}^{N} \omega_{ij} \boldsymbol{K}_j^X(\mathcal{H}) \right) \boldsymbol{A}'.
\end{align}

By using the multiplicative property of the determinant and $\det(\boldsymbol{A}) \neq 0$ we see that $\boldsymbol{A}$ is only a constant in the minimization problem and is not affecting the choice of the optimal combination of subsets,
\begin{align*}
    f(\mathcal{H}) =& \sum_{i = 1}^{N}  \det\left( (1-\lambda) \boldsymbol{K}_i^Y(\mathcal{H}) + \lambda \sum_{j = 1 , j \neq i}^{N} \omega_{ij} \boldsymbol{K}_j^Y(\mathcal{H}) \right) \\
    =& \det(\boldsymbol{A})^2 \sum_{i = 1}^{N}  \det\left( (1-\lambda) \boldsymbol{K}_i^X(\mathcal{H}) + \lambda \sum_{j = 1 , j \neq i}^{N} \omega_{ij} \boldsymbol{K}_j^X(\mathcal{H}) \right).
\end{align*}
Together with Equation~(\ref{eq:proof_affineequi}), affine equivariance is proven for the covariance estimator. Since the location estimator is defined as the arithmetic mean, which is affine equivariant, 
the property stated in Equation~(\ref{eq:def_affineequivariant}) is also fulfilled.  
\end{proof}
\proofinappendixstatement

The theorem assumes that the estimator $\boldsymbol{T}(\boldsymbol{X})$ is regular, robust and affine equivariant which are properties that can be achieved by the MCD for the full data set $\boldsymbol{X}$ for $n > 2p$. 
However, this is a non-trivial problem in cases where we do not have enough observations or where additional regularization would be needed. Since this is the situation where the MRCD should provide a remedy, the target matrix for the original MRCD will most likely not fulfill the assumptions stated in the theorem. Thus, the MRCD is in general not affine equivariant.

Another important property of robust estimators is the finite sample breakdown point, 
which is defined as the minimal fraction of points that need to be exchanged in order to make the estimators useless. 
Before considering the spatially smoothed MRCD we have to derive the breakdown point of the original MRCD without prior scaling of the observations, from now on called \textit{raw MRCD}. 
The breakdown point of a location estimator $\hat{\boldsymbol{\mu}}_{n}$ is formally defined as
\begin{align*}
    \epsilon_{n}^* (\hat{\boldsymbol{\mu}}_{n}; \boldsymbol{X}_{n}) = \frac{1}{n} \min \{ m: \sup || \hat{\boldsymbol{\mu}}_{n}(\boldsymbol{X}_{n,m}) - \hat{\boldsymbol{\mu}}_{n}(\boldsymbol{X}_n) || = + \infty \},
\end{align*}
where $\boldsymbol{X}_{n,m}$ is the data matrix $\boldsymbol{X}_n$ with $m$-many observations exchanged with arbitrary values \citep{maronna06}.

For the covariance estimate $\hat{\boldsymbol{\Sigma}}_{n}$ the finite sample breakdown point is defined as
\begin{align*}
    \epsilon_{n}^* (\hat{\boldsymbol{\Sigma}}_{n}; \boldsymbol{X}_{n}) = \frac{1}{n}\min \{ m:  \sup \max_j | \ln(\lambda_j(\hat{\boldsymbol{\Sigma}}_{n}(\boldsymbol{X}_{n,m}))) -\ln(\lambda_j( \hat{\boldsymbol{\Sigma}}_{n}(\boldsymbol{X}_n))) | = + \infty \},
\end{align*}
with $\lambda_1(\boldsymbol{\Sigma}), \ldots ,\lambda_p(\boldsymbol{\Sigma})$ denoting the eigenvalues of a matrix $\boldsymbol{\Sigma}$ in decreasing order. Since the eigenvalues are sorted, we only have to consider the biggest eigenvalue $\lambda_1(\hat{\boldsymbol{\Sigma}}_{n}(\boldsymbol{X}_{n,m})))$ which might explode when exchanging observations with arbitrary values (\textit{explosion breakdown point}) and the smallest eigenvalue $\lambda_p(\hat{\boldsymbol{\Sigma}}_{n}(\boldsymbol{X}_{n,m})))$ which might become zero (\textit{implosion breakdown point}) and thus implies singularity \citep{maronna06}.

\begin{theoremrep}
\label{theorem_MRCD_breakdown}
Consider the raw MRCD estimator with fixed $\rho > 0$, regular and fixed $\boldsymbol{T} = \boldsymbol{Q\Lambda Q}'$ and the data matrix $\boldsymbol{X}_n$. Then, the following statements hold:
\begin{itemize}
    \item [a.]The MRCD location estimator $\hat{\boldsymbol{\mu}}_{n}$ has the finite sample breakdown point $\min(h, n-h+1)/n$.
    \item [b.]The MRCD covariance estimator $\hat{\boldsymbol{\Sigma}}_{n}$ has the finite sample explosion breakdown point $(n-h+1)/n$.
    \item [c.]The MRCD covariance estimator $\hat{\boldsymbol{\Sigma}}_{n}$ has the finite sample implosion breakdown point $1$.
\end{itemize}
\end{theoremrep}

\begin{proof}
\textbf{c.} Regarding notation see also \cite{Boudt2020}. 
The eigenvalues for the transformed covariance matrix $\boldsymbol{K_W} = \rho \boldsymbol{I} + (1-\rho)c_{\alpha} \boldsymbol{S_W}(H)$, where $\boldsymbol{S_W}(H)$ is the covariance matrix of a subset $H$ of $\boldsymbol{X}$, are bounded below by $\rho > 0$.
Thus, $\lambda_i(\boldsymbol{K_W}^{-1}) = \lambda_i(\boldsymbol{K_W})^{-1} \le 1/\rho$ and $\norm{\boldsymbol{K_W}^{-1}  }_2 \le 1/\rho$ where $\norm{.}_2$ denotes the spectral matrix norm. For covariance matrices the spectral norm is equal to its biggest eigenvalue. 
It follows that
\begin{align*}
    \norm{ \left(\boldsymbol{Q \Lambda}^{1/2} \boldsymbol{K_W}  \boldsymbol{\Lambda}^{1/2} \boldsymbol{Q}' \right)^{-1} }_2 &= 
    \norm{ \boldsymbol{Q \Lambda}^{-1/2} \boldsymbol{K_W}^{-1}  \boldsymbol{\Lambda}^{-1/2} \boldsymbol{Q}' }_2 \\
    &\le \norm{ \boldsymbol{Q' \Lambda}^{-1/2} }_2^2 \norm{\boldsymbol{K_W}^{-1}}_2 \\
    & \le c/\rho,
\end{align*}
for $c>0$, since $\boldsymbol{Q' \Lambda}^{-1/2}$ is also regular and fixed.
This implies that 
\begin{align*}
    \lambda_i\left(\boldsymbol{Q \Lambda}^{1/2} \boldsymbol{K_W}  \boldsymbol{\Lambda}^{1/2} \boldsymbol{Q}'\right)^{-1} = \lambda_i\left((\boldsymbol{Q \Lambda}^{1/2} \boldsymbol{K_W}  \boldsymbol{\Lambda}^{1/2} \boldsymbol{Q}' )^{-1} \right) \le c/\rho 
\end{align*}
for all $i = 1, \ldots, p$ .
It follows that 
\begin{align*}
    \lambda_i(\boldsymbol{Q \Lambda}^{1/2} \boldsymbol{K_W}  \boldsymbol{\Lambda}^{1/2} \boldsymbol{Q}') \ge \rho/c > 0  \ \ \forall i = 1, \ldots, p,
\end{align*}
for all subsets $H$, specifically for the optimal subset $H^*$. Thus, the eigenvalues of $\hat{\boldsymbol{\Sigma}}_{n} = \boldsymbol{Q \Lambda}^{1/2} \boldsymbol{K_W}^*  \boldsymbol{\Lambda}^{1/2} \boldsymbol{Q}'$ are also bounded away from zero, and it follows that the implosion breakdown point is 1.
\hfill

\medskip
\textbf{b.} 
It is clear that  $\epsilon_{n}^* (\hat{\boldsymbol{\Sigma}}_{n}; \boldsymbol{X}_{n}) \le (n-h+1)/n$ since in this case there would always be at least one observation in the selected subset independent of its value spoiling the estimation. 
We need to show that $\epsilon_{n}^* (\hat{\boldsymbol{\Sigma}}_{n}; \boldsymbol{X}_{n}) > (n-h)/n$.

Suppose  $\epsilon_{n}^* (\hat{\boldsymbol{\Sigma}}_{n}; \boldsymbol{X}_{n}) \le (n-h)/n$. 
We can change $m \le n-h$ observations arbitrarily and denote the resulting matrix as $\boldsymbol{X}_{n,m} = (\boldsymbol{x}_1^*, \ldots,  \boldsymbol{x}_m^*, \boldsymbol{x}_{m+1}, \ldots, \boldsymbol{x}_{n})'$, where $\boldsymbol{x}_1^*, \ldots, \boldsymbol{x}_m^*$ are the exchanged observations (w.l.o.g. placed in the first $m$ rows). 
The supremum being infinite is equivalent to
\begin{align}
\label{eq:MRCD1}
    \forall C>0 \ \exists  \boldsymbol{X}_{n,m}: \ \ | \ln(\lambda_1(\hat{\boldsymbol{\Sigma}}_{n}(\boldsymbol{X}_{n,m}))) -\ln(\lambda_1( \hat{\boldsymbol{\Sigma}}_{n}(\boldsymbol{X}_n))) | > C.
\end{align}
Since $\ln(\lambda_1 (\hat{\boldsymbol{\Sigma}}_{n}(\boldsymbol{X}_n))$ is constant and $\ln$ is monotonously increasing and unrestricted we can w.l.o.g. assume that the value inside of the absolute value is non-negative. 
Additionally, moving the constant $\ln(\lambda_1( \hat{\boldsymbol{\Sigma}}_{n}(\boldsymbol{X}_n)))$ to the right hand side, Equation~\eqref{eq:MRCD1} is equivalent to the unboundedness of the biggest eigenvalue $\lambda_1(\hat{\boldsymbol{\Sigma}}_{n}(\boldsymbol{X}_{n,m}))$, 
\begin{align}
\label{eq:MRCD2}
     \forall C>0 \ \exists ~ \boldsymbol{x}_1^*, \ldots, \boldsymbol{x}_m^*: \ \  \lambda_1(\hat{\boldsymbol{\Sigma}}_{n}(\boldsymbol{X}_{n,m})) > C.
\end{align}
Note that $\det(\boldsymbol{X}) = \prod_{i =1}^p \lambda_i( \boldsymbol{X})$ for any $p$-dimensional matrix $ \boldsymbol{X}$ and that  $\hat{\boldsymbol{\Sigma}}_{n}(\boldsymbol{X}_{n,m}) = (1-\rho)c_{\alpha}Cov(\boldsymbol{X}_{n,m; H^*}) + \rho \boldsymbol{T}$, where $H^*$ is the subset of observations of $\boldsymbol{X}_{n,m}$ that minimize $\det((1-\rho)c_{\alpha}Cov(\boldsymbol{X}_{n,m; H}) + \rho \boldsymbol{T})$ over all subsets $H$.
As shown above, the eigenvalues of $(1-\rho)c_{\alpha} Cov(\boldsymbol{X}) + \rho \boldsymbol{T}$ are bounded away from zero for all $\boldsymbol{X}$ and $\rho > 0$,
\begin{align*}
\lambda_i( (1-\rho)c_{\alpha}Cov(\boldsymbol{X}) + \rho \boldsymbol{T}) \ge c > 0, \ \ \forall i = 1, \ldots, p.
\end{align*}
For the matrix $\boldsymbol{X}_{n,m; H^*}$, it follows that
\begin{align*}
    \prod_{i =2}^p \lambda_i(\hat{\boldsymbol{\Sigma}}_{n}(\boldsymbol{X}_{n,m})) \ge c^{p-1} > 0.
\end{align*}

Let the constant $\tilde{C}$ be defined as
\begin{align*}
   \tilde{C}  = \frac{\det((1-\rho)c_{\alpha}Cov(\boldsymbol{X}_{n,m; \tilde{H}}) + \rho \boldsymbol{T})}{c^{p-1}},
\end{align*}
where $\tilde{H} = {m+1, \ldots, m+h}$ denote $h$ indices of fixed and unchanged observations of $\boldsymbol{X}_{n,m}$, which exist due to $m\le n-h$. 
Then, due to condition~\eqref{eq:MRCD2} for $\tilde{C}$ there exists $\boldsymbol{x}_1^*, \ldots , \boldsymbol{x}_m^*$ such that $\lambda_1(\hat{\boldsymbol{\Sigma}}_{n}(\boldsymbol{X}_{n,m})) > \tilde{C}$ which leads to 
\begin{align*}
    \det((1-\rho)c_{\alpha}Cov(\boldsymbol{X}_{n,m; H^*}) + \rho \boldsymbol{T}) > \det((1-\rho)c_{\alpha}Cov(\boldsymbol{X}_{n,m; \tilde{H}}) + \rho \boldsymbol{T}).
\end{align*}
This contradicts the minimization of the determinant property of the selected subset $ H^*$. Thus, $\epsilon_{n}^* (\hat{\boldsymbol{\Sigma}}_{n}; \boldsymbol{X}_{n}) > (n-h)/n$.

\medskip
\textbf{a.}
Using the same argument as before, it is clear that  $\epsilon_{n}^* (\hat{\boldsymbol{\mu}}_{n}; \boldsymbol{X}_{n}) \le (n-h+1)/n$.
Let us show that  $\epsilon_{n}^* (\hat{\boldsymbol{\mu}}_{n}; \boldsymbol{X}_{n}) \le h/n$. 
Again we can argue that $\sup || \hat{\boldsymbol{\mu}}_{n}(\boldsymbol{X}_{n,m}) - \hat{\boldsymbol{\mu}}_{n}(\boldsymbol{X}_n) || = + \infty$ is equivalent to 
\begin{align}
\label{eq:MRCD3}
   \forall C>0 \ \exists ~ \boldsymbol{x}_1^*, \ldots, \boldsymbol{x}_m^*: \ \norm{ \hat{\boldsymbol{\mu}}_{n}(\boldsymbol{X}_{n,m})} > C.
\end{align}
We have to find $m = h$ many exchanged data points in a way that the norm of the location estimator is unbounded but the determinant of the covariance matrix is still minimal.
For the fixed data set $\boldsymbol{X}_n$, we obtain the optimal subset $H^*$ of observations and add a fixed but arbitrarily large number $L>0$ to the first coordinate of these $m = h$ observations,
\begin{align*}
    \forall i \in H^*: \tilde{x}_{i1} = x_{i1} + L \text{ and } \tilde{x}_{ij} = x_{ij} \ \forall j = 2, \ldots, p.
\end{align*}
Thus, the sample mean of the first coordinate of
the selected subset $\boldsymbol{X}_{n,m;H^*}$ is equal to the original 
mean of the first coordinate of $\boldsymbol{X}_{n;H^*}$ plus $L$. Similarly, the sample covariance is the same as before given that we take the same subset $H^*$, since it is independent of constant shifts applied to all used observations. This implies that also the regularized covariance and its determinant are the same which was minimal for all other subsets of $\boldsymbol{X}_{n}$. In order to show minimality of the subset $H^*$ for the new data matrix $\boldsymbol{X}_{n,m}$ it follows that we only have to consider the subsets that have both original and exchanged (arbitrarily large) observations. 

Regarding the sample mean of the first coordinate of one of these subsets $\tilde{H}$, it is 
\begin{align*}
    \tilde{M}_1 = \frac{1}{h} \left(\sum_{j= 1}^k x_{i_j1} + \sum_{j = k+1}^h \tilde{x}_{i_j1} \right) = \frac{1}{h} \left(\sum_{j= 1}^k x_{i_j1} + \sum_{j = k+1}^h (x_{i_j1} + L) \right) = 
    M_1 + L - \frac{k+1}{h}L,
\end{align*}
where $M_1$ is the (fixed) mean of the first coordinate of the subset $\boldsymbol{X}_{n;\tilde{H}}$ and both sums are not empty. Regarding the variance of the first coordinate, which is the first diagonal entry of the sample covariance matrix, we see
\begin{align*}
    Cov(\boldsymbol{X}_{n,m; \tilde{H}})_{11} &= \frac{1}{h-1} \left(\sum_{j= 1}^k (x_{i_j1} - \tilde{M}_1)^2 + \sum_{j = k+1}^h (\tilde{x}_{i_j1} - \tilde{M}_1)^2 \right) \\
    &= \frac{1}{h-1} \left(\sum_{j= 1}^k \underbrace{(x_{i_j1} - M_1 - L + \frac{k+1}{h}L)^2}_{\mathcal{O}(L^2)} + \sum_{j = k+1}^h \underbrace{(x_{i_j1} + L - M_1 - L + \frac{k+1}{h}L)^2}_{\mathcal{O}(L^2)} \right)  \\
    &= \mathcal{O}(L^2)
\end{align*}
Thus, the Frobenius norm of $Cov(\boldsymbol{X}_{n,m; \tilde{H}})$ and also its regularization are $\mathcal{O}(L^2)$ and it follows for some constant $\beta > 0$ that 
\begin{align*}
    \mathcal{O}(L^2) &= \norm{(1-\rho)c_{\alpha}Cov(\boldsymbol{X}_{n,m; \tilde{H}}) + \rho \boldsymbol{T})}_F \\
    &\le \beta \norm{(1-\rho)c_{\alpha}Cov(\boldsymbol{X}_{n,m; \tilde{H}}) + \rho \boldsymbol{T})}_2 \\
    &= \beta\lambda_1((1-\rho)c_{\alpha}Cov(\boldsymbol{X}_{n,m; \tilde{H}}) + \rho \boldsymbol{T}) \\
    &\le \beta\frac{\det((1-\rho)c_{\alpha}Cov(\boldsymbol{X}_{n,m; \tilde{H}})}{c^{p-1}},
\end{align*}
due to equivalence of matrix norms in finite dimensional space and $c$ being the constant from above. Choosing $L$ arbitrarily large, we see that the determinant corresponding to a mixed subset is larger than the determinant of the optimal subset $H^*$ of only exchanged observations. 

Now suppose $\epsilon_{n}^* (\hat{\boldsymbol{\mu}}_{n}; \boldsymbol{X}_{n}) = m/n < \min(h, n-h+1)/n$ and start from Equation~\eqref{eq:MRCD3},
\begin{align*}
    \forall C>0 \ \exists ~ \boldsymbol{x}_1^*, \ldots, \boldsymbol{x}_m^*: \norm{ \hat{\boldsymbol{\mu}}_{n}(\boldsymbol{X}_{n,m}) } > C.
\end{align*}
This implies that 
\begin{align*}
    \forall C>0 \ \exists ~ \boldsymbol{x}_1^*, \ldots, \boldsymbol{x}_m^*: \sum_{j= 1}^k \norm{\boldsymbol{x}_{i_j}} + \sum_{j = k+1}^h \norm{\boldsymbol{x}^*_{i_j}}  \ge \norm{ \left(\sum_{j= 1}^k 
    \boldsymbol{x}_{i_j} + \sum_{j = k+1}^h \boldsymbol{x}^*_{i_j} \right) } > C,
\end{align*}
where $i_j \in H^*, j = 1, \ldots, h$. Thus, for all $C>0$ 
there exists some $\boldsymbol{x}^*_{i_j}$ whose norm is bigger than $C$. W.l.o.g. assume it is $\boldsymbol{x}_1^*$ and that the first coordinate is responsible,
\begin{align*}
    \forall C>0 \ \exists ~ \boldsymbol{x}_1^* \in H^*:  |x^*_{1 1}|  \ge C.
\end{align*}

For $m < h < n-h+1$ there would not be the possibility to only include exchanged points in the subset and it would always be possible to have a subset of $h$ many original observations. This is also the case for $m < n-h+1 < h$. Thus, there are at least one exchanged point $\boldsymbol{x}_1^*$ and one original point in $H^*$. But with the same argument as before, the determinant of the mixed subset of original points and arbitrarily large points would eventually contradict optimality, because at one point the determinant would be so large that there would be an $h$-sized subset of original observations available to get a smaller determinant. 
\end{proof}
\proofinappendixstatement

Regarding the finite sample breakdown point of location and covariance estimators with multiple estimators let us define the finite sample breakdown point $\epsilon_{n}^*$ as the minimal fraction of points of the same arbitrary neighborhood that need to be exchanged in order to make at least one of the estimators useless. For the location estimators $\hat{\boldsymbol{\mu}}_{SSM, n, i}$, $ i = 1, \ldots, N$, the formal definition is 
\begin{equation*}
    \epsilon_{n}^* (\left(\hat{\boldsymbol{\mu}}_{SSM, n, i}\right)_{i = 1}^N; \boldsymbol{X}_{n}) = \min_{i = 1, \ldots, N} \frac{1}{n_i} \min \{ m: \sup || \hat{\boldsymbol{\mu}}_{SSM, n, i}(\boldsymbol{X}_{n,m}^i) - \hat{\boldsymbol{\mu}}_{SSM, n, i}(\boldsymbol{X}_n) || = + \infty \},
\end{equation*}
where $\boldsymbol{X}_{n,m}^i$ is the matrix $\boldsymbol{X}_n$ with $m$-many observations of neighborhood $a_i$ exchanged with arbitrary values. 

\begin{theoremrep}
    The location estimators $\left(\hat{\boldsymbol{\mu}}_{SSM, n, i}\right)_{i = 1}^N$ of the spatially smoothed MRCD have the finite sample breakdown point 
    \begin{align*}
        \epsilon_{n}^* (\left(\hat{\boldsymbol{\mu}}_{SSM, n, i}\right)_{i = 1}^N; \boldsymbol{X}_{n}) = \min_{i=1, \ldots, N} \min(n_i-h_i+1, h_i)/n_i.
    \end{align*}
\end{theoremrep} 

\begin{proof}
    For $i =1, \ldots, N$, the location estimate is the standard sample mean of $h_i$ many observations from neighborhood $a_i$ selected in a way to minimize the objective function~\eqref{eq:objective_function}. By exchanging observations in one neighborhood $a_i$ with arbitrarily large values and keeping the other neighborhoods the same (keeping the matrices $\boldsymbol{K}_j$ bounded), we can apply the results of Theorem~\ref{theorem_MRCD_breakdown} for the MRCD structured covariance matrix $\boldsymbol{K}_i$. The breakdown point for $\boldsymbol{K}_i$ is $\min(n_i-h_i+1, h_i)/n_i$. Thus, in order to make at least one of the location estimators useless we need to exchange a fraction $\min_{i=1, \ldots, N} \min(n_i-h_i+1, h_i)/n_i$ of observations of one neighborhood.
\end{proof}
\proofinappendixstatement

\medskip
For the covariance estimates $\hat{\boldsymbol{\Sigma}}_{SSM, n,i}$, $ i = 1, \ldots, N$, the finite sample breakdown point is defined accordingly,
\begin{align*}
    \epsilon_{n}^* (\left(\hat{\boldsymbol{\Sigma}}_{SSM, n,i}\right)_{i = 1}^N; \boldsymbol{X}_{n}) &= \min_{i = 1, \ldots, N} \frac{1}{n_i}\min \{ m: \sup \max_j | \ln(\lambda_j(\hat{\boldsymbol{\Sigma}}_{SSM, n,i}(\boldsymbol{X}_{n,m}^i))) -\ln(\lambda_j( \hat{\boldsymbol{\Sigma}}_{SSM, n,i}(\boldsymbol{X}_n))) | = + \infty \}.
\end{align*}
Again, we can differentiate between explosion and implosion breakdown point.

\begin{theoremrep}
Given a fixed and regular target matrix $\boldsymbol{T}$, the finite sample implosion breakdown point of the spatially smoothed MRCD covariance estimators $\left(\hat{\boldsymbol{\Sigma}}_{SSM, n,i}\right)_{i=1}^N$ is equal to 
\begin{align*}
    \epsilon_{n}^* (\left(\hat{\boldsymbol{\Sigma}}_{SSM, n,i}\right)_{i = 1}^N; \boldsymbol{X}_{n}) = 1.
\end{align*}
The finite sample explosion breakdown point is
\begin{align*}
    \epsilon_{n}^* (\left(\hat{\boldsymbol{\Sigma}}_{SSM, n,i}\right)_{i = 1}^N; \boldsymbol{X}_{n}) = \min_{i=1, \ldots, N} (n_i-h_i+1)/n_i.
\end{align*}
\end{theoremrep}

\begin{proof}
Since the spatially smoothed MRCD covariance estimators $\boldsymbol{K}_i$ are regularized on each neighborhood according to the MRCD approach, all eigenvalues are positive and bounded away from zero as long as the target matrix $\boldsymbol{T}$ is regular (see Theorem~\ref{theorem_MRCD_breakdown}). Hence, none of the covariance estimators will ever be singular and the implosion breakdown point is 1.

For the second part let us fix neighborhood $a_i$. 
Note, that the covariance estimator is defined as $\hat{\boldsymbol{\Sigma}}_{SSM, n,i} = (1-\lambda) \boldsymbol{K}_i(\mathcal{H}^*) + \lambda \sum_{j = 1 , j \neq i}^{N} \omega_{ij} \boldsymbol{K}_j(\mathcal{H}^*)$ for the optimal subset $\mathcal{H}^*$. 
The matrix $\boldsymbol{K}_i = \boldsymbol{K}_i(\mathcal{H}^*)$ is structured in an MRCD manner based on the sample covariance matrix of the subset $H_i^*$ and the target matrix. 
Since we assume $\boldsymbol{T}$ to be fixed, for $\boldsymbol{K}_i$ we can get arbitrarily large eigenvalues only under the same circumstances as for the MRCD (see Theorem~\ref{theorem_MRCD_breakdown}). Thus, exchanging a fraction of $(n_i-h_i+1)/n_i$ by arbitrary values can lead to arbitrarily large eigenvalues of $\boldsymbol{K}_i$. 
For the explosion breakdown point for one neighborhood covariance estimator $\hat{\boldsymbol{\Sigma}}_{SSM, n,i}$ (under general settings for $\boldsymbol{W}$ and $\lambda$) it is sufficient that at least one $\boldsymbol{K}_j$ has reached its breakdown point. It follows, that the finite sample explosion breakdown point of $\hat{\boldsymbol{\Sigma}}_{SSM, n,i}$ is 
\begin{align}
\label{eq:explosion_BP}
    \epsilon_{n}^* (\hat{\boldsymbol{\Sigma}}_{SSM, n,i}; \boldsymbol{X}_{n}) = \min_{i=1, \ldots, N} \{(n_i-h_i+1)/n_i\}.
\end{align}
Since $\epsilon_{n}^* (\hat{\boldsymbol{\Sigma}}_{SSM, n,i}; \boldsymbol{X}_{n})$ is already independent of $i$, the overall explosion breakdown point for the spatially smoothed MRCD covariance estimators is equal to $\epsilon_{n}^* (\hat{\boldsymbol{\Sigma}}_{SSM, n,i}; \boldsymbol{X}_{n})$.
\end{proof}
\proofinappendixstatement

Note that for all the breakdown properties of the original and the spatially smoothed MRCD, the target matrix $\boldsymbol{T}$ is assumed to be regular and fixed. In applications the target matrix would be some covariance estimator $\boldsymbol{T}(\boldsymbol{X})$. Thus, it should be kept in mind that the implosion and explosion breakdown point of $\boldsymbol{T}(.)$ might also be relevant.

\subsection{Local outlier detection}
The final step for detecting outliers is to decide how the spatially smoothed covariances available for neighborhoods $a_i$, $i = 1, \ldots, N$, will be linked to the pairwise MD. 

The method is based on \cite{Ernst2016}. In order to compare each observation $\boldsymbol{x}$ with its local neighbors we need a second neighborhood structure that provides spatially close neighbors in contrast to the structural neighborhoods $a_i$ that are used for the smoothed covariance estimation. Thus, we select some $k \in \mathbb{N}$ and calculate the spatial $k$ nearest neighbors, $b_k(\boldsymbol{x})$, see Equation~\eqref{eq:spatially_knn_def}, where a 
typical value might be $k = 10$. However, when applying the method, the spatial structure of the data set should also be considered. 

For each observation $\boldsymbol{x} \in a_i$, the \textit{next distance} is defined as
\begin{align*}
    d(\boldsymbol{x}) = min_{\boldsymbol{y} \in b_k(\boldsymbol{x})} \left[ (\boldsymbol{x} - \boldsymbol{y})' \hat{\boldsymbol{\Sigma}}_{SSM,i} ^{-1} (\boldsymbol{x} - \boldsymbol{y}) \right] ^{1/2},
\end{align*}
which is the minimum of all pairwise MDs based on the covariance matrix $\hat{\boldsymbol{\Sigma}}_{SSM,i}$ with all observations in the spatial $k$ nearest neighborhood $b_k(\boldsymbol{x})$. The neighborhood $b_k(\boldsymbol{x})$ is not necessarily a subset of $a_i$. However, due to the spatial smoothing of the covariance matrices and the spatial correlation of regular observations, an observation $\boldsymbol{y}$ spatially close to $\boldsymbol{x}$ should be similar enough to not be classified as outlier even if the covariance matrix of another but close neighborhood $a_i$ is used. In the case of a strong difference between $\boldsymbol{x}$ and $\boldsymbol{y}$, the observation $\boldsymbol{y}$ would still be classified as outlying.

The next distance $d(\boldsymbol{x})$ is used as a measure of outlyingness. If the next distance is high,
none of the observations in the spatial $k$ nearest neighborhood is similar to the observation $\boldsymbol{x}$, which means that
$\boldsymbol{x}$ would be flagged as a local outlier. 
As a notion what values for the next distance are considered as high, a non-parametric cut-off value can be used based on the upper fence of the adjusted boxplot \citep{Hubert2008} using all next distances from all neighborhoods $a_i$, $i = 1, \ldots, N$. Observations above the cut-off value are considered to be local outliers.

Possible further extensions like the restriction to homogeneous neighborhoods as suggested in \cite{Ernst2016} are not included but are interest of future research.

\section{Algorithm and C-step}
\label{sec:algorithm}

For computing the spatially smoothed MRCD location and covariance estimators
we need to minimize the objective function~(\ref{eq:objective_function}).
However, since the number of possible combinations of subsets is comparable with the MCD or the MRCD, it is in general not feasible to just 
calculate the value of the objective function for all these combinations 
and select the best one.
Instead, the strategy of C-steps (concentration steps) introduced
for the MCD estimator by \cite{Rousseeuw1999} will be adapted to this problem setting. 
Given an index set $H_1$ corresponding to $h$ observations
of the data matrix $\boldsymbol{X}$, the C-step chooses the subsequent subset $H_2$ where the $h$ observations with the smallest Mahalanobis distances, based on the arithmetic mean and sample
covariance matrix of the observations from $H_1$, are taken. 
\cite{Rousseeuw1999} have shown that this procedure converges to a local minimum.
The C-step idea has also been extended for the MRCD \citep{Boudt2020}, and we will now adapt the generalized C-step to our setting.

\begin{theoremrep}
\label{theorem_Cstep}
For each $j = 1, \ldots , N$, let $\rho_j \in (0, 1)$ and $H_j^0$ be any starting subset of $a_j$ of respective size $h_j \in (n_j/2, n_j)$. Let $\alpha_j$ be the corresponding fraction of the observations used, $\alpha_j = h_j/n_j$.
Let $\mathcal{H}^0=(H_1^0, \ldots, H_N^0)$ be the combination of the subsets and $\lambda \in [0,1)$ fixed. 
The target matrix $\boldsymbol{T}(\boldsymbol{X})$ is assumed to be positive definite, symmetric and fixed, and $\boldsymbol{K}_j(\mathcal{H}^0)$ is defined as in Equation~(\ref{eq:K_def}) with $\boldsymbol{T} = \boldsymbol{T}(\boldsymbol{X})$. 
Calculate the sample mean for each neighborhood $a_j, j = 1, \ldots, N$, based on the respective subset, $\boldsymbol{m}_{j}(\mathcal{H}^0) = \frac{1}{h_j} \sum_{k \in H_{j}^0} \boldsymbol{x}_k$.

Fix neighborhood $a_i$. For $\boldsymbol{x} \in a_i$, let the MD-based measure with the subset given by $\mathcal{H}^0$ be defined as
\begin{align*}
	d(\boldsymbol{x}; \mathcal{H}^0) = (\boldsymbol{x}-\boldsymbol{m}_{i}(\mathcal{H}^0))' \left[ (1-\lambda) \boldsymbol{K}_i({\mathcal{H}^0})+ \lambda \sum_{j = 1, j \neq i}^N \omega_{ij} \boldsymbol{K}_{j}(\mathcal{H}^0)  \right]^{-1}  (\boldsymbol{x}-\boldsymbol{m}_{i}(\mathcal{H}^0)).
\end{align*}
For a new subset $H_i^1$ of $a_i$ of size $h_i$ with 
\begin{align*}
	\sum_{k \in H_i^1} d(\boldsymbol{x}_k; \mathcal{H}^{0}) \leq \sum_{k \in H_i^0} d(\boldsymbol{x}_k; \mathcal{H}^0),
\end{align*}
denote $\tilde{\mathcal{H}} = (H_1^0, \ldots, H_i^1, \ldots , H_N^0)$ (note that $\boldsymbol{K}_{j}(\tilde{\mathcal{H}}) = \boldsymbol{K}_{j}(\mathcal{H}^0)$ for $j \ne i$).
Then, 
\begin{align*}
    \det \left( (1-\lambda) \boldsymbol{K}_i(\tilde{\mathcal{H}})+ \lambda \sum_{j = 1, j \neq i}^N \omega_{ij} \boldsymbol{K}_{j}(\tilde{\mathcal{H}}) \right) \leq \det\left((1-\lambda) \boldsymbol{K}_i({\mathcal{H}^0})+ \lambda \sum_{j = 1, j \neq i}^N \omega_{ij} \boldsymbol{K}_{j}(\mathcal{H}^0)  \right) 
\end{align*}
with equality if and only if $\boldsymbol{K}_i(\tilde{\mathcal{H}}) = \boldsymbol{K}_i({\mathcal{H}^0})$ and $\boldsymbol{m}_{i}(\tilde{\mathcal{H}}) = \boldsymbol{m}_{i}(\mathcal{H}^0)$.
\end{theoremrep}

\begin{proof}
This proof is very much along the lines of \cite{Boudt2020}.
The neighborhood $a_i$ is fixed. 
Thus, the matrix which determinant should be minimized regarding $i$ is 
\begin{align*}
	\boldsymbol{A}_1 & := \left((1-\lambda) \boldsymbol{K}_i({\mathcal{H}^0})+ \lambda \sum_{j = 1, j \neq i}^N \omega_{ij} \boldsymbol{K}_{j}(\mathcal{H}^0) \right) \\
	& = \left( (1-\lambda) [(1-\rho_i) c_{\alpha_i} Cov(\boldsymbol{X}_{H_i^0}) + \rho_i \boldsymbol{T}] + \lambda \sum_{j = 1, j \neq i}^N \omega_{ij} \boldsymbol{K}_{j}(\mathcal{H}^0)  \right)  \\
	& = \left(\underbrace{(1-\lambda) (1-\rho_i)c_{\alpha_i}}_{:=\tilde{\rho}} Cov(\boldsymbol{X}_{H_i^0}) + \underbrace{(1-\lambda)\rho_i \boldsymbol{T} + \lambda \sum_{j = 1, j \neq i}^N \omega_{ij} \boldsymbol{K}_{j}(\mathcal{H}^0) }_{:= \boldsymbol{\Omega}} \right) \\
	& = \tilde{\rho} ~ Cov(\boldsymbol{X}_{H_i^0})  + \boldsymbol{\Omega}, \\
	\boldsymbol{A}_2 &: = \left((1-\lambda) \boldsymbol{K}_i({\tilde{\mathcal{H}}})+ \lambda \sum_{j = 1, j \neq i}^N \omega_{ij} \boldsymbol{K}_{j}(\tilde{\mathcal{H}}) \right) \\
	& = \tilde{\rho} ~ Cov(\boldsymbol{X}_{H_i^1})  + \boldsymbol{\Omega},
\end{align*}
where $\boldsymbol{\Omega}$ is a fixed positive definite covariance matrix. 

Since the original proof is not restricted to convex linear combinations, we can use the same proof with the matrices $\boldsymbol{A}_1$, $\boldsymbol{A}_2$ and $\boldsymbol{\Omega}$ in place of $\boldsymbol{K}_1$, $\boldsymbol{K}_2$ and $\boldsymbol{\Lambda}$ and adapted factors
\begin{align*}
	w_j &= \sqrt{k \tilde{\rho} / h_i }, & & \ j = 1, \ldots, h_i \\
	& = \sqrt{k / (p+1) }, & & \ j = h_i + 1, \ldots, k.
\end{align*}
with $k = h_i + p + 1$.\footnote{Note that this proof can be generalized to any kind of linear combination with fixed matrices and a sample covariance matrix of a subset.}
\end{proof}
\proofinappendixstatement
	
\medskip	
The C-step theorem states that the objective function will decrease with every step as long as the other covariance estimators stay fixed. 
In the implemented algorithm described below, this will in general not be the case. 
However, the theorem and its proof should be sufficient to motivate and provide a reason for the algorithm proposed in the following. 

The algorithm makes use of the C-step to solve the minimization problem based on \cite{Boudt2020}.
Suppose that we can estimate the target matrix $\boldsymbol{T} = \boldsymbol{T}(\boldsymbol{X})$ by the affine equivariant MCD estimator, then we also obtain affine equivariance for the spatially smoothed MRCD. Thus, we can ignore the scaling step in \cite{Boudt2020} and reduce the number of parameter estimations. Using an eigen-decomposition of $\boldsymbol{T} = \boldsymbol{Q} \boldsymbol{\Lambda} \boldsymbol{Q}'$, with $\boldsymbol{Q}$ containing the eigenvectors as columns, and 
$\boldsymbol{\Lambda}$ being the diagonal matrix of positive eigenvalues, we transform the observations, 
\begin{equation}
\label{eq:transform}
    \boldsymbol{z}_i = \boldsymbol{\Lambda}^{-1/2} \boldsymbol{Q}' \boldsymbol{x}_i ,
\end{equation}
for $i=1,\ldots ,n$. Thus, in the next steps we can use $\boldsymbol{Z} = (\boldsymbol{z}_1, \ldots, \boldsymbol{z}_n)'$ as data matrix and $\boldsymbol{T} = \boldsymbol{I}_p$ as target matrix, which numerically simplifies the data-driven selection procedure for $\rho_{j}$ .

In order to start the iteration process by making use of the C-steps, we need starting values, which should be robust and regular covariance estimates for each neighborhood. 
As proposed for the original MRCD estimator, we will also use the deterministic MCD algorithm of \cite{Hubert2012} for each neighborhood separately. 
This approach results in six estimates for location and scatter for each neighborhood, which show especially good convergence properties. 
Furthermore, for each neighborhood $a_j$, the value of $\rho_{j}$ is calculated using the data-driven selection procedure based on a maximal condition number according to steps 3.2 to 3.4 from \cite{Boudt2020}. 
The set of six deterministic starting values for each neighborhood is restricted to those with a sufficiently small condition number \citep[for more details see step 3.5. in][]{Boudt2020}.

One new issue that arises is the number of possible combinations of initial subsets:
for $N$ neighborhoods we end up with up to $6^N$ subset combinations as possible starting values. 
Since a high number of starting values might not be essential for a good approximation,
we will consider an upper limit of $6N$ starting values in the following, and they will
be selected at random out of the possible combinations that are left after the $\rho$-selection step.

\begin{figure}[htp]
    \centering
    
    \begin{tikzpicture} 
        \node(Start)[anchor=west] at (0,4) {$Starting \ values \ H_i^0 \ \ \forall i = 1, \ldots, N$ given};
        
        \node(00)[anchor=west] at (0,3) {$Step\ 0:$};
        \node(H01) at (3,3) {$H_1^0$};
        \node(H02) at (5,3) {$H_2^0$};
        \node(H0x) at (7,3) {$\ldots$};
        \node(H03) at (9,3) {$H_N^0$};
        
        \node(01) at (3,2) {$\boldsymbol{K}_1^0$};
        \node(02) at (5,2) {$\boldsymbol{K}_2^0$};
        \node(0x) at (7,2) {$\ldots$};
        \node(03) at (9,2) {$\boldsymbol{K}_N^0$};
        
        \node(H10)[anchor=west]  at (0,-1) {$Step\ 1:$};
        \node(H11) at (3,-1) {$H_1^1$};
        \node(H12) at (5,-1) {$H_2^1$};
        \node(H1x) at (7,-1) {$\ldots$};
        \node(H13) at (9,-1) {$H_N^1$};
        
        \node(11) at (3,-2) {$\boldsymbol{K}_1^1$};
        \node(12) at (5,-2) {$\boldsymbol{K}_2^1$};
        \node(1x) at (7,-2) {$\ldots$};
        \node(13) at (9,-2) {$\boldsymbol{K}_N^1$};
        
        \node(22) at (6,-3) {$\vdots$};
        
        \node(end)[anchor=west]  at (0,-4.2) {$Until \ convergence: \ H_i^m = H_i^{m-1} \ \ \forall i = 1, \ldots, N$};
        \node(cstep1) at (8.3, 0.6) {\textit{C-step}};
        \node(cstep1) at (8.3, -3.2) {\textit{C-step}};

        \draw(H01)[-stealth] to (01);
        \draw(H02)[-stealth] to (02);
        \draw(H03)[-stealth] to (03);
        
        \draw [decorate, decoration = {brace,mirror, raise=0.35cm}] (2.75,2) --  (9.25,2);
        
        \draw(6,1.4)[-stealth] to (H11);
        \draw(6,1.4)[-stealth] to (H12);
        \draw(6,1.4)[-stealth] to (H13);
        \draw(6,1.4)[-stealth] to (7,-0.7);
        
        \draw(H11)[-stealth] to (11);
        \draw(H12)[-stealth] to (12);
        \draw(H13)[-stealth] to (13);
        
        \draw [decorate, decoration = {brace,mirror, raise=0.35cm}] (2.75,-2) --  (9.25,-2);

    \end{tikzpicture}
    \caption{Matrices used in the C-step after each iteration step. $H_i^j$ are the selected subsets of neighborhood $a_j$ in step $i$, and $\boldsymbol{K}_i^j$ the corresponding regular covariance matrices.}
    \label{fig:iteration_procedure}
\end{figure}
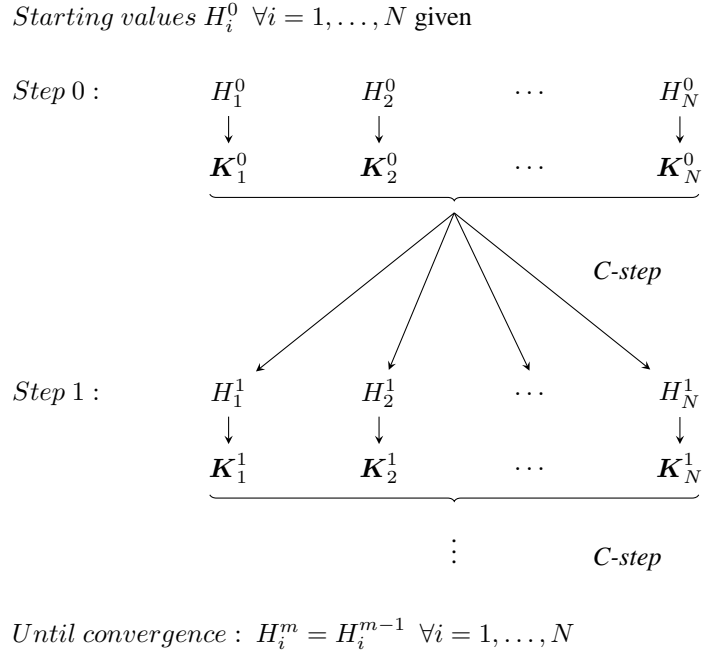

Suppose now that we start the procedure with an initial subset 
$\mathcal{H}^0 =  (H_1^0, \ldots, H_N^0)$, then we apply the C-step for each neighborhood 
$a_i$ and obtain a new subset $H_1^i$. The combination of these subsets 
$\mathcal{H}^1 =  (H_1^1, \ldots, H_N^1)$ is used as starting point for the next iteration step, etc.
After there is no change in the subsets, the iteration process stops (see Figure~\ref{fig:iteration_procedure}).
After applying the C-step iterations for all starting values, we choose the subset combination with the smallest objective function value as the final subset combination $\mathcal{H}^* = (H_1^*, \ldots, H_N^*)$.

Although Theorem~\ref{theorem_Cstep} is not proving that the objective function decreases with every step due to the additional covariance matrices being adapted separately for each neighborhood after each iteration step, simulation results show that the algorithm provides stable results and good monotonic behavior in most cases (see Section~\ref{sec:Numerical_simulations}).

After receiving the final combination of subsets $\mathcal{H}^*$ for each neighborhood, the matrices $\boldsymbol{K}_i^Z$ are back-transformed to
\begin{align}
\label{eq:K_backtransform}
	\boldsymbol{K}_i^* = \boldsymbol{Q} \boldsymbol{\Lambda} ^{1/2} \left[ \boldsymbol{K}_i^Z (\mathcal{H}^*)\right] \boldsymbol{\Lambda} ^{1/2} \boldsymbol{Q}'.
\end{align}
The covariance estimate for neighborhood $a_i$ is then
\begin{align}
\label{eq:final_estimate}
    \hat{\boldsymbol{\Sigma}}_{SSM,i} = (1-\lambda) \boldsymbol{K}_{i}^*+ \lambda \sum_{j = 1, j \neq i}^N \omega_{ij} \boldsymbol{K}_{j}^*,
\end{align}
and the location estimate is the arithmetic mean of the optimal subset $\boldsymbol{X}_{\mathcal{H}^*_i}$.
The detailed numerical procedure is summarized as pseudo-code in 
Algorithm~1.

Regarding the tuning parameter $\lambda$ there is no standard procedure to get a good value that is similarly automated like the calculation procedure for the regularization parameters $\rho_i$. However, there are multiple possibilities demonstrated in Section~\ref{sec:examples} that can be used to choose $\lambda$ in applications.

	\small
	\begin{algorithm}

        {\bf Step 1.1:} Calculation of target matrix $\boldsymbol{T}$ using the MCD estimator on $\boldsymbol{X}$\;
		{\bf Step 1.2:} Eigen-decomposition of $\boldsymbol{T} = \boldsymbol{Q} \boldsymbol{\Lambda} \boldsymbol{Q}'$ and transform observations according to Equation~(\ref{eq:transform})\;
		{\bf Step 2:} Initialization step according to steps 3.1 to 3.5 (without C-step) from \cite{Boudt2020} for each neighborhood\;
		\For{$i= 1, \ldots ,N$} {
		    Fix neighborhood $a_i$\;
			Get 6 initial deterministic sets of $h_i$ observations from $a_i$ according to \cite{Hubert2012}\;
			Calculate 6 initial covariance matrices and mean estimates\;
			Select neighborhood specific $\rho_i$ via data driven approach\;
			Filter subset of initial starting estimates according to condition number \citep[step 3.5 from][]{Boudt2020}\;
			}
		Select set of initial h-set combinations as starting values at random\;
		{\bf Step 3.1:} C-step: For each initial combination of subsets iterate until convergence\;
		{\bf Step 3.2:} Select best combination of subset based on objective function value\;
		{\bf Step 4:} Calculate $\boldsymbol{K}_i(\mathcal{H}^*)~ \forall i$\ and use Equations~\eqref{eq:K_backtransform}~and~\eqref{eq:final_estimate} to get final estimates\; 
		\caption{Algorithm for the spatially smoothed MRCD estimator.}
		\label{alg:general}
	\end{algorithm}
	\normalsize

\pagebreak
\section{Numerical simulations}
\label{sec:Numerical_simulations}

In order to test the new method, two simulation scenarios are set up which also incorporate the neighborhood structures necessary for the covariance estimation and outlier detection. For both setups we 
simulate covariance matrices which depend on a parameter $\delta$, denoted by $\boldsymbol{\Sigma}\left(\delta\right)$, where the entries are defined as $\left(\boldsymbol{\Sigma}\left(\delta\right)\right)_{jk} = \delta^{(j-k)}$, for $j,k \in \{1,\ldots ,p\}$,
leading to positive definiteness and symmetry.

\textit{Setup 1:} The first setup is inspired by the original idea of covariance matrices smoothly transforming over space. We start by setting up the (two-dimensional) coordinates $(s^1_i, s^2_i)$ of the observations $\boldsymbol{x}_i$ with $n_{sim} = 41$ observations per coordinate axis, evenly spread between $0$ and $20$, resulting in $n = n_{sim}^2 = 1681$ data points in total. 

For $N_{sim}$ many areas $a_{lm}$, $l,m \in \{1, \ldots ,\sqrt{N_{sim}}\}$, of similar observations we construct a second spatial grid with each cell consisting of $n_{sim}/N_{sim}$ observations on average. The borders of the areas for the first coordinate are defined as $b_l^1 = l \frac{20}{\sqrt{N_{sim}}}$ for $l = 0, \ldots, \sqrt{N_{sim}}$. Analogously, the borders for the second coordinate are defined as $b_m^2 =  m \frac{20}{\sqrt{N_{sim}}}$ for $m = 0, \ldots, \sqrt{N_{sim}}$, leading the an evenly spaced grid. Thus, the area $a_{lm}$ consists of observations $\{\boldsymbol{x}_i: s^1_i \in (b_{l-1}^1, b_l^1] \cap s^2_i \in (b_{m-1}^2, b_m^2] \}$ for $l, m \neq 0$. For either $l$ or $m$ equal to $1$, the left edge of the interval (which would be zero) is included. The coordinate centers of the area $a_{lm}$ are defined as $c_{lm}^1 = \frac{b_{l-1}^1 + b_l^1}{2}$ and $c_{lm}^2 =\frac{b_{m-1}^2 +b_m^2}{2}$. 

\begin{figure}[htp]
    \centering
    \includegraphics[width = 0.49\textwidth]{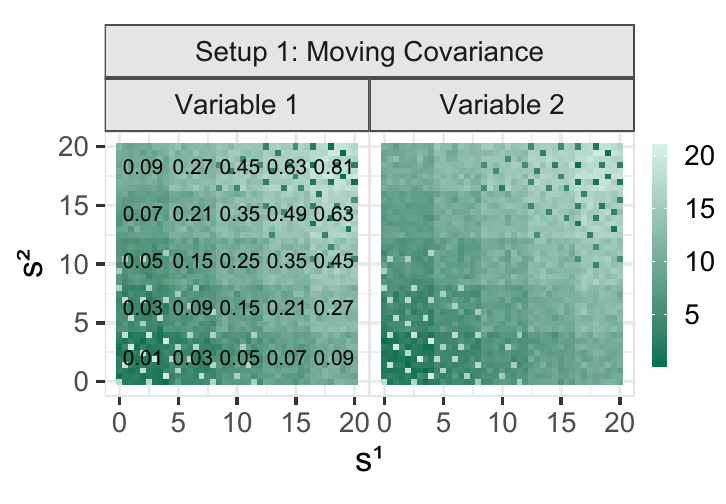}
    \includegraphics[width = 0.49\textwidth]{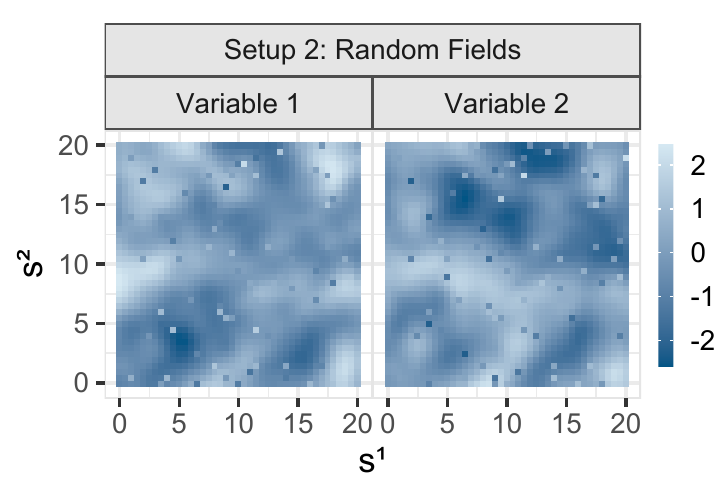}
    \caption{Simulation scenarios with $p=2$ and a $5\%$ contamination rate. On the left hand side the simulation setup 1 is presented with contamination achieved through the swapping process described in \cite{Ernst2016}, $N_{sim} = 25$ and $n_{sim} = 41$. The values printed on the left-most panel are corresponding to the parameter $\delta_{lm}$. On the right hand side, setup 2 with $\nu = 3$ is shown with completely random swapping.}
    \label{fig:simulationDesign}
\end{figure}

The observed values of observations in area $a_{lm}$ are then randomly drawn from a $p$-dimensional normal distribution $\mathcal{N}\left( \boldsymbol{\mu}_{lm}, \boldsymbol{\Sigma}\left(\delta_{lm}\right)  \right)$, where $\boldsymbol{\mu}_{lm} := ((c_{lm} ^1 + c_{lm} ^2)/2, \ldots, (c_{lm} ^1 + c_{lm} ^2)/2) \in \mathbb{R}^{p \times 1}$, thus having entry values between $0$ and $20$. 
For the covariance matrix $\boldsymbol{\Sigma}\left(\delta_{lm}\right)$ for areas $a_{lm}$ we use the structure described above with parameter $\delta_{lm}$ defined as
\begin{equation*}
    \delta_{lm} = \left( 0.1 + l \frac{0.9-0.1}{\sqrt{N_{sim}}} \right) \left( 0.1 + m \frac{0.9-0.1}{\sqrt{N_{sim}}} \right) \in [0.01, 0.81],
\end{equation*}
increasing smoothly from the left bottom to the right upper corner. A simulated data set with $p=2$ is presented in Figure~\ref{fig:simulationDesign} (left) where the grid structure and the change of the mean are clearly visible. The resulting $\delta_{lm}$ for each area is shown as well.

\textit{Setup 2:} To get a more flexible simulation setup, a random field, specifically the parsimonious multivariate Mat\'ern model \citep[see][]{Gneiting2010, Ernst2016}, is used as suggested in \cite{Ernst2016} and \cite{Harris2013}. Instead of the constructed matrices used in \cite{Ernst2016} and \cite{Harris2013} we again choose a matrix structure $\boldsymbol{\Sigma}\left(\delta\right)$ as described above with $\delta = 0.7$, since it can be extended for higher dimensions. For $\delta$ being set to $0.7$ we get a range of high to low correlations reflecting different relationships between variables. The spatial smoothness is assumed to be the same for all variables, and it is regulated by one smoothness parameter $\nu$, which is taking values in $\{0.5, 1.5, 3\}$. A higher $\nu$ leads to more spatial smoothness and in general more distinct outliers after contamination. The spatial scale parameter $a$ of the Mat\'ern model is set to one. A grid structure for the coordinates of the observations with values ranging from $0$ to $20$ with grid size $0.5$ is imposed, leading to $41^2 = 1681$ observations overall, similar to the standard setting in setup 1. 

Lastly, contamination with outliers is achieved by swapping coordinates of observations with each other. \cite{Filzmoser2013} exchange observations that are completely randomly chosen, whereas \cite{Ernst2016} propose to swap the most extreme observations regarding the first score of the global robust principal components. In order to avoid the problem of exchanging whole areas of observations with each other due to high spatial correlation, once observations are swapped, their $15$ closest neighbors are removed from the swapping process. This leads to a clear distinction of outlying observations without the possibility of other outliers being close (see also Figure~\ref{fig:simulationDesign}). Thus, swapping according to \cite{Ernst2016} should in general result in a better performance for all considered methods. Both swapping approaches in both setups will be analyzed with a varying contamination level $\beta$ between $1\%$ and $10\%$.

For the ssMRCD covariance estimation we impose a grid based neighborhood structure. Similar to the description of setup 1 we use evenly spaced borders and assign the observations to neighborhoods $n_i$, for $i=1, \ldots, N$. Thus, the case $N = N_{sim}$ in setup 1 depicts a perfect match of the neighborhoods selected for the ssMRCD and the real underlying covariance structure. For $N_{sim} > N$ the ssMRCD uses less neighborhoods leading to more smoothness of the covariance estimation. The weighting matrix $\boldsymbol{W}$ is based on the inverse (Euclidean) distance of the centers, that are defined equivalently to setup 1.

\subsection{Covariance estimation}

\begin{figure}[htp]
    \centering
    \includegraphics[width = \textwidth]{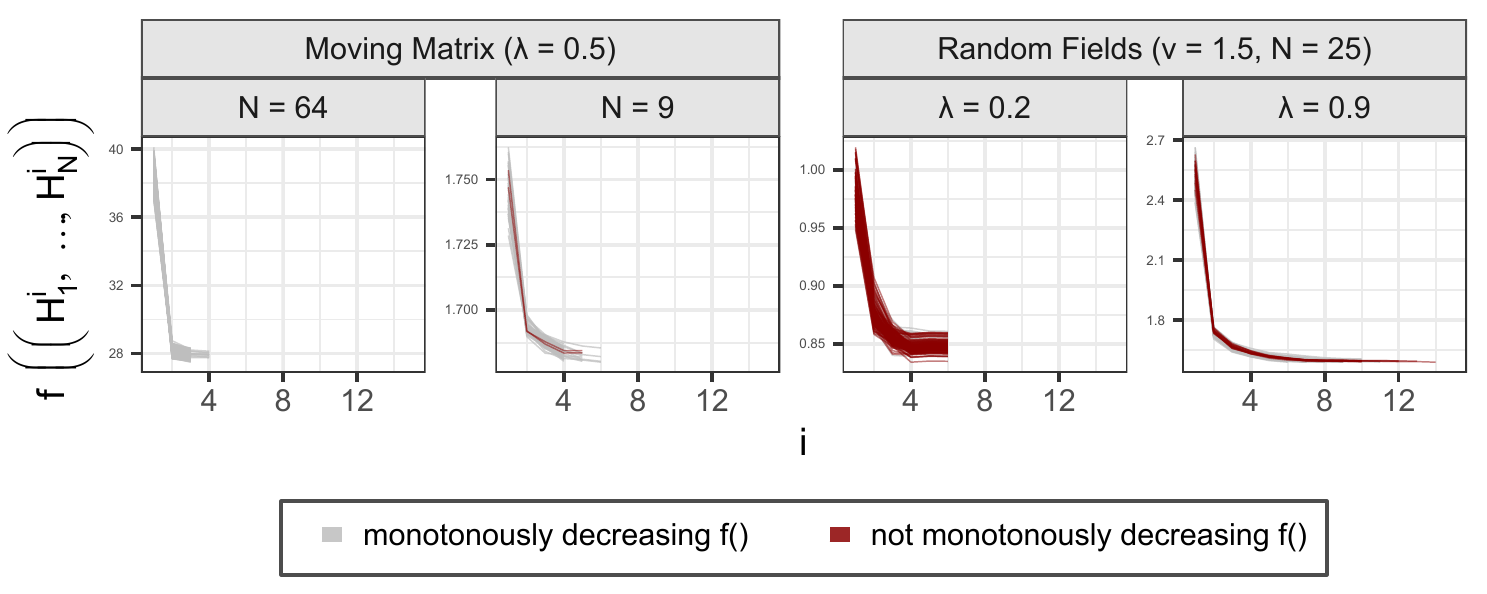}
    \caption{Different convergence behavior of objective function. A $p = 5$ dimensional setting with $5\%$ contamination achieved with completely random swapping. Each line is the path of one initial set of H-sets along the C-step iteration in the algorithm in Section~\ref{sec:algorithm}.}
    \label{fig:convergence}
\end{figure}

First, we want to analyse the algorithm described and motivated in Section~\ref{sec:algorithm} in more detail. Since Theorem~\ref{theorem_Cstep} is only valid for one varying covariance matrix and not for multiple varying ones, the convergence properties should be further examined. Figure~\ref{fig:convergence} shows the objective function values along the iteration procedure for all starting H-set combinations for different parameter settings for both simulation setups. We choose $p = 5$ and a $5\%$ contamination rate achieved by completely random swapping. The weighting matrix is based on inverse distances as mentioned in Section~\ref{sec:method}.
Each panel reflects the convergence behavior of the objective function for one simulated data set.

Although the behavior differs in general, it is evident that the algorithm has overall very good convergence properties. A high percentage of monotonically decreasing objective functions 
can often be achieved and the non-monotonically decreasing paths increase only marginally. 
The reliability of the convergence results in the simulation might possibly be due to the spatially correlated values which lead to rather small changes in the covariance matrices during the algorithm. Thus, for our simulated data sets, the assumption of fixed covariance matrices seems to be sufficiently met. 
Since in reality local outlier detection mostly makes sense only for spatially correlated data, the theoretical results from Theorem~\ref{theorem_Cstep} proof to be even more valuable.  

Moreover, the convergence takes place fast which might be caused by the choice of the good starting estimates of the detMCD algorithm \citep{Hubert2012}. Another reason might be that the number of observations that can be used is restricted by the neighborhood assignments to a smaller number than in a covariance estimation for the full data set. Very promising is also the rapid improvement at the very beginning, independent of the simulated data sets.

A second issue that should be discussed is computational efficiency. The iteration process and the increasing number of starting values with increasing $N$ can have quite an impact on runtime. Nevertheless, as long as the number of neighborhoods $N$ is not too big, the outlier detection method based on the ssMRCD is competitive with other local methods \citep{Ernst2016}, especially if $p$ is large (see also Figure~\ref{fig:runtime_comparison}. For simulation setup~1 with $N_{sim} = 25$, a $5\%$ contamination rate through completely random switching and $100$ repetitions are used and the parameters $p$, $\lambda$, $N$ and $n$ are each varied univariately. The default values for parameters not being varied are $p = 5$, $\lambda = 0.5$, $N = 25$, and $n = 1681$.

\begin{figure}[htp]
    \centering
    \includegraphics[width = \textwidth]{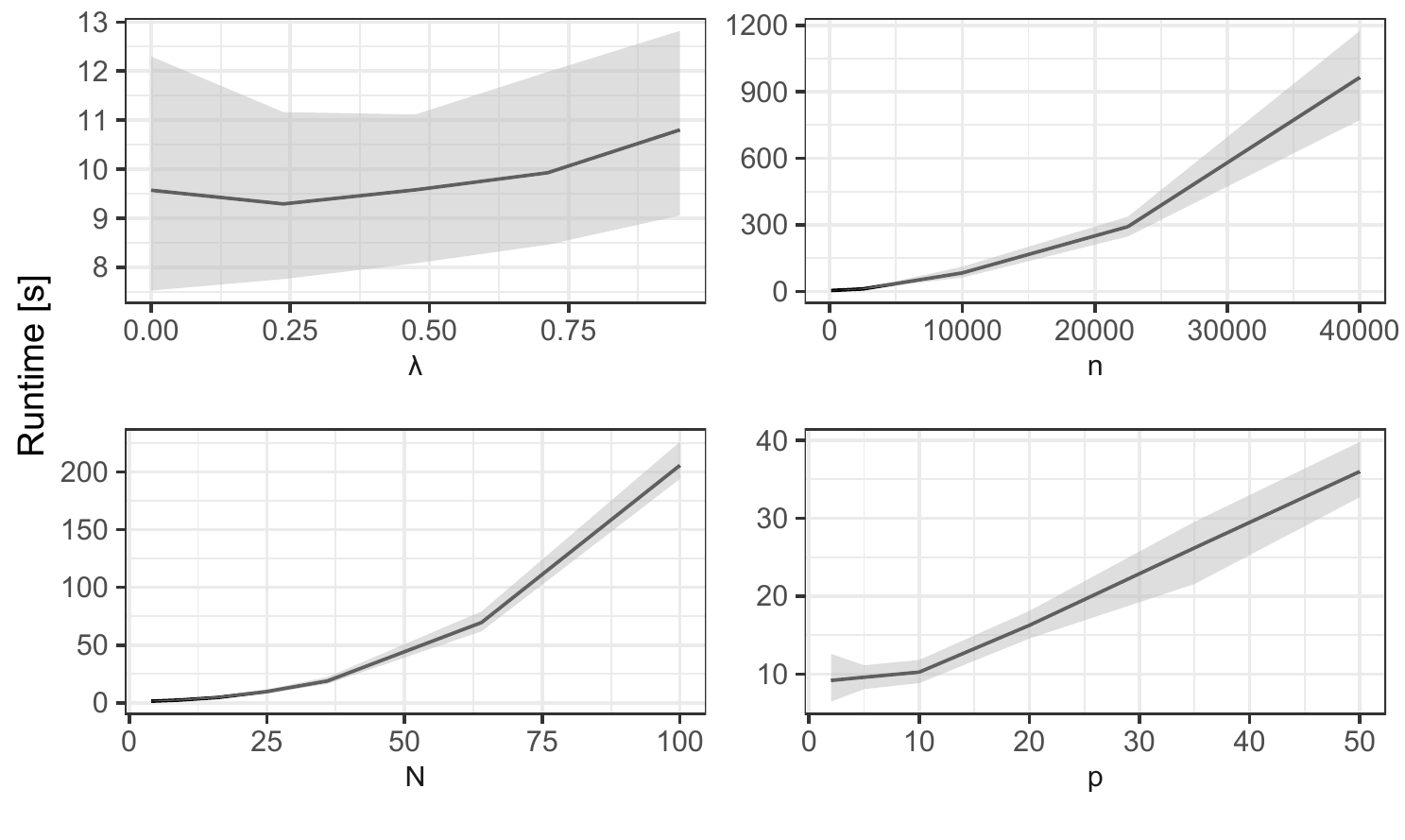}
    \caption{Analysis of runtime of the ssMRCD for setup 1 with $5\%$ contamination rate, $N_{sim} = 25$ and varying parameter values with default values $\lambda = 0.5$, $p = 5$, $N = 25$ and $n = 1681$ if not varied. The solid line is representing the mean of $100$ repetitions, the edges of the gray band around the mean the $5\%$ and $95\%$ quantile.}
    \label{fig:runtime}
\end{figure}

As depicted in Figure~\ref{fig:runtime}, the number of neighborhoods $N$ and the number of observations $n$ have the most influence on runtime with more than a linear increase. The nearly quadratic increase for $N$ is partly due to the number of starting values increasing linearly with $N$, which could also be reduced to enhance efficiency if necessary. Interestingly, the dimension $p$ has an approximately linear effect on runtime which is overall moderate. The smoothing parameter $\lambda$ does not significantly change the runtime. 

\subsection{Outlier detection}

Before comparing the performance in local outlier detection with other methods, the parameter sensitivity of the ssMRCD is analyzed in more detail. This simulation study should simplify the choices of $\lambda$ and $N$ in particular for real world data and focus on possible issues connected to suboptimal parameter settings.

For this purpose, setup~2 (random fields) is considered as simulation setting, with parameter  $\nu = 3$, and $\beta = 5\%$ completely randomly swapped observations. Special focus is put on the choice of $\lambda$ and $N$, but also the effect of dimension $p$ is analyzed. The other parameters are chosen in accordance to possible default settings. The weighting matrix is based on the inverse Euclidean distances of the centers of the neighborhoods $a_i$. Since all considered methods propose $k = 10$ as a default value, we adhere to this setting for now. Each parameter combination was simulated for $100$ different realizations. While \cite{Ernst2016} suggest to use Cohen's Kappa as summary statistic of the confusion matrix, we will use the F1-score due to its good interpretability and suitability also for imbalanced classification data.

\begin{figure}
    \centering
    \includegraphics[width = \textwidth]{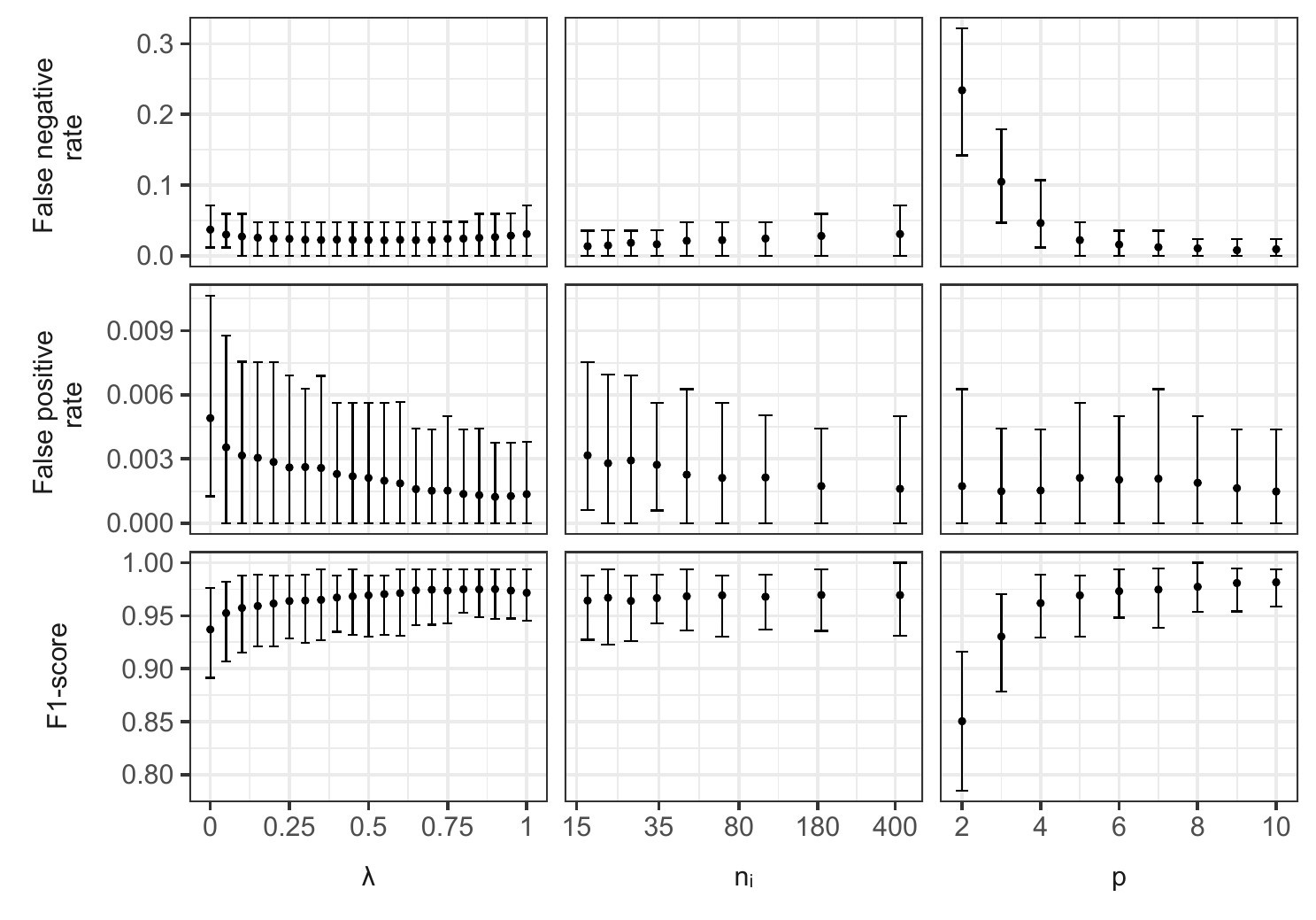}
    \caption{Outlier detection performance based on the false negative, false positive rate and the F1-score of the ssMRCD outlier detection method for different parameter settings. Each point represents the arithmetic mean and the corresponding bars the 5th and 95th quantile of 100 simulations. For non-varying parameters the default settings are $p = 5$, $N = 25$ (comparable to $n_i \approx 67$) and $\lambda = 0.5$.}
    \label{fig:parameter_sensitivity}
\end{figure}

Figure \ref{fig:parameter_sensitivity} shows the false negative rate, the false positive rate and the resulting F1-score plotted against varying values of $\lambda$, $N$ and $p$, with default values of $p = 5$, $N = 25$ (implying $n_i \approx 67$ on average) and $\lambda = 0.5$. These values should reflect a quite general and unspecific parameter setting. For illustration purposes it is more informative to plot the average neighborhood size $n_i \approx 1681/N$ instead of the number of neighborhoods.

The simulation results show that a higher $\lambda$ decreases the false positive rate, it has marginal reduction effects on the false negative rate until too much smoothing masks real outliers. The overall performance increases moderately in $\lambda$, but for $\lambda$ higher than $0.5$ the increase is marginal. Thus, we propose a default value of $0.5$ for $\lambda$ to get the advantage of the decrease in the false positive rate while avoiding the masking effect for higher values. Compared to the influence of $\lambda$, the effect of the dimension $p$ is more pronounced. Very small dimensions seem to cover outliers more effectively, probably due to less available information. Interestingly, the size of the neighborhoods seems to be relatively irrelevant, at least in this simulation setting. Only a small masking effect occurs similar to the effects of $\lambda$. Too big neighborhoods lead to too much smoothing. Thus, this might imply to choose a strategy of medium sized neighborhoods to increase efficiency in computation and reduce unnecessary regularization.
This guidance for the parameter choices might be biased towards this specific simulation setting and not optimal in other settings, but fixing the parameters with at least a sensible value simplifies the overall procedure.

Regarding the suitability of the ssMRCD covariances for local outlier detection we compare its performance of outlier detection to the local outlier detection methods of \cite{Filzmoser2013} (F), \cite{Ernst2016}(EH) and the local outlier factor methodology of \cite{Breunig2000}, canonically adapted to spatial neighborhoods as described in \cite{Schubert2012} (LOF). Both simulation setups vary in the parameters $p$, $\nu$ and $N_{sim}$, respectively, and both swapping processes are used, each combination simulated $100$ times. All methods considered compare each observation to $k$ many of its neighboring observations which we will assume to be equal to $10$ for all methods. For the ssMRCD we will assume the default values ($\lambda = 0.5$, $N = 25$, and $\boldsymbol{W}$ based on inverse-distances) from the prior setup for parameter sensitivity analysis. 

The outlier classification method of \cite{Filzmoser2013} has a parameter $\beta_F$ which is the percentage of neighboring observations that a local outlier is allowed to be similar to. Here we use the value $\beta_F=0.1$, as proposed in \cite{Ernst2016}, meaning that $0.1k = 1$ observation is allowed to be similar within the $10$ nearest neighbors. For inliers the expected value of the isolation degree is $\beta_F$. If the actual degree of isolation is higher than the expected value, this signals local outlyingness. As cutoff for classifying an observation as local outlier we use twice the value of $\beta_F$, so $20\%$. This cutoff is less strict than in the simulation setup from \cite{Ernst2016} who take a cutoff of three times the expected value, i.e.~$30\%$. Note that the methodology of \cite{Filzmoser2013} mostly focuses on visual outlier detection tools, so the cutoff value chosen here might not be optimal.

For the regularized spatial detection technique by \cite{Ernst2016}, the parameter $\beta_{EH}$, which gives the fraction of the most homogeneous neighborhoods included in the outlier detection procedure, is set to one. In the simulations we are interested in all outliers, and the heterogeneity in the simulated data should be comparable for all of the observations. Thus, only considering a fraction leads to non comparable results. Moreover, the simulation results of \cite{Ernst2016} show that over all considered setups, $\beta_{EH} = 1$ is also optimal. As regular covariance matrix estimator we use the MRCD with the default target matrix (equi-correlated target matrix) and $\alpha = 75\%$.

Last but not least, the non-parametric LOF, which calculates a local outlier factor for each observation based on a comparison of the so-called "local reachability density" with its k-nearest neighbors, needs a cutoff value. Since there is no fixed rule on how to choose a cutoff value, a local outlier factor above $1.5$ determines an outlier. This value is also used in the original paper of \cite{Breunig2000}.

\begin{figure}[htp]
    \centering
    \includegraphics[width = 0.95\textwidth, trim = {0 0.6cm 0 0}, clip  ]{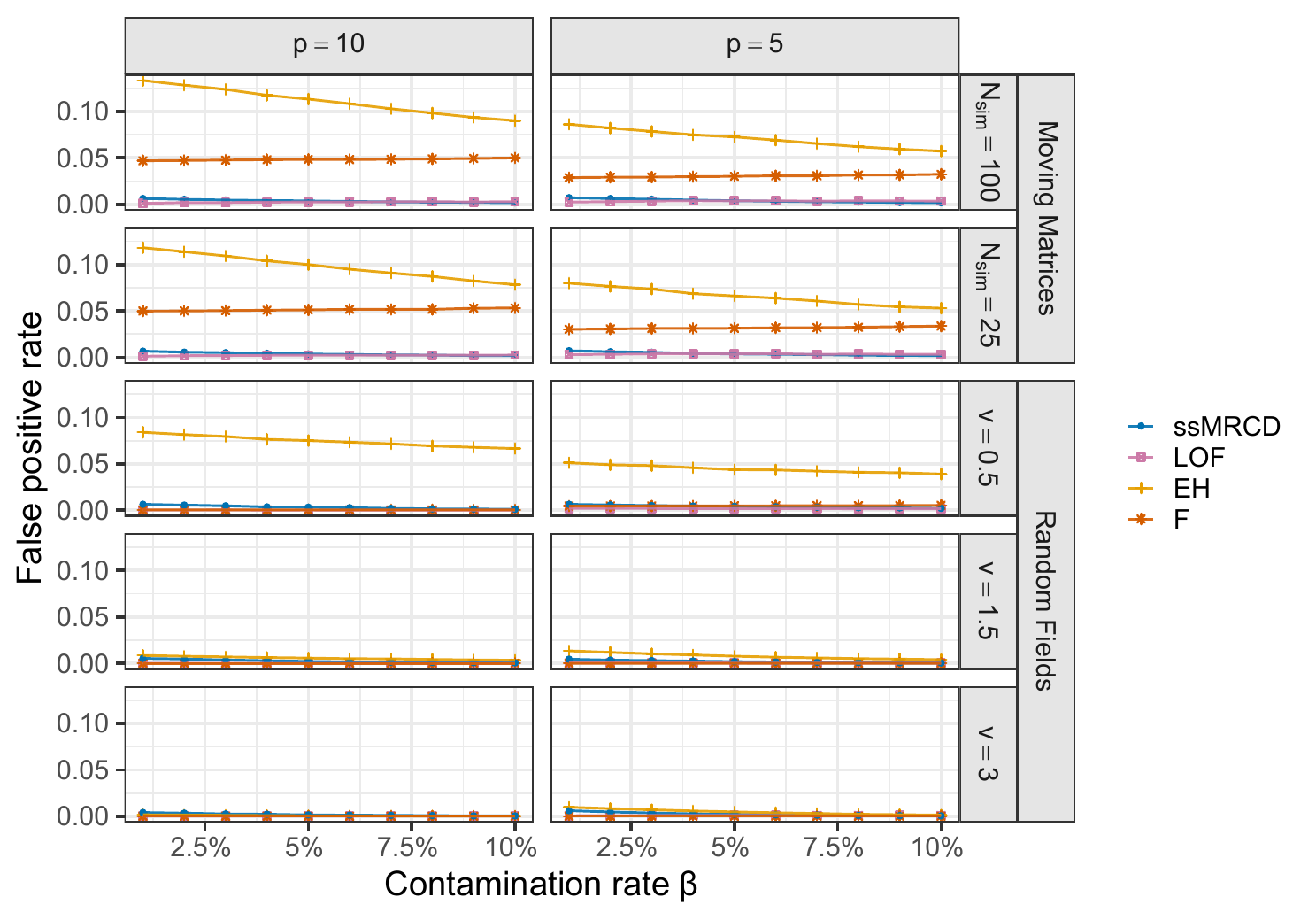}
    \includegraphics[width = 0.95\textwidth]{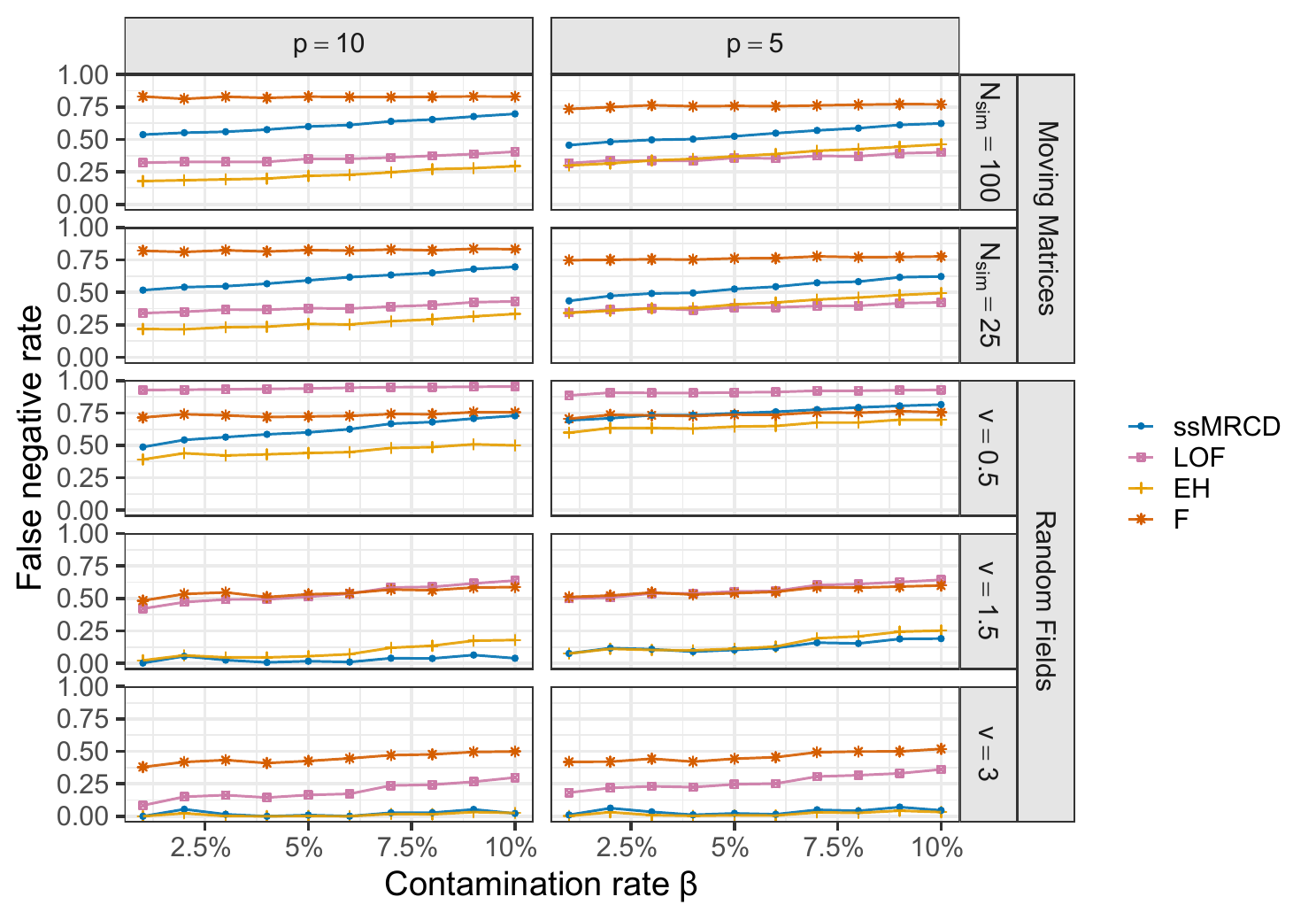}
    \caption{False positive and false negative rate for all four outlier detection methods with contamination achieved through completely random switching for different scenarios. Each point represents the mean of 100 repetitions.}
    \label{fig:Harris1}
\end{figure}

\begin{figure}[htp]
    \centering
    \includegraphics[width = 0.95\textwidth]{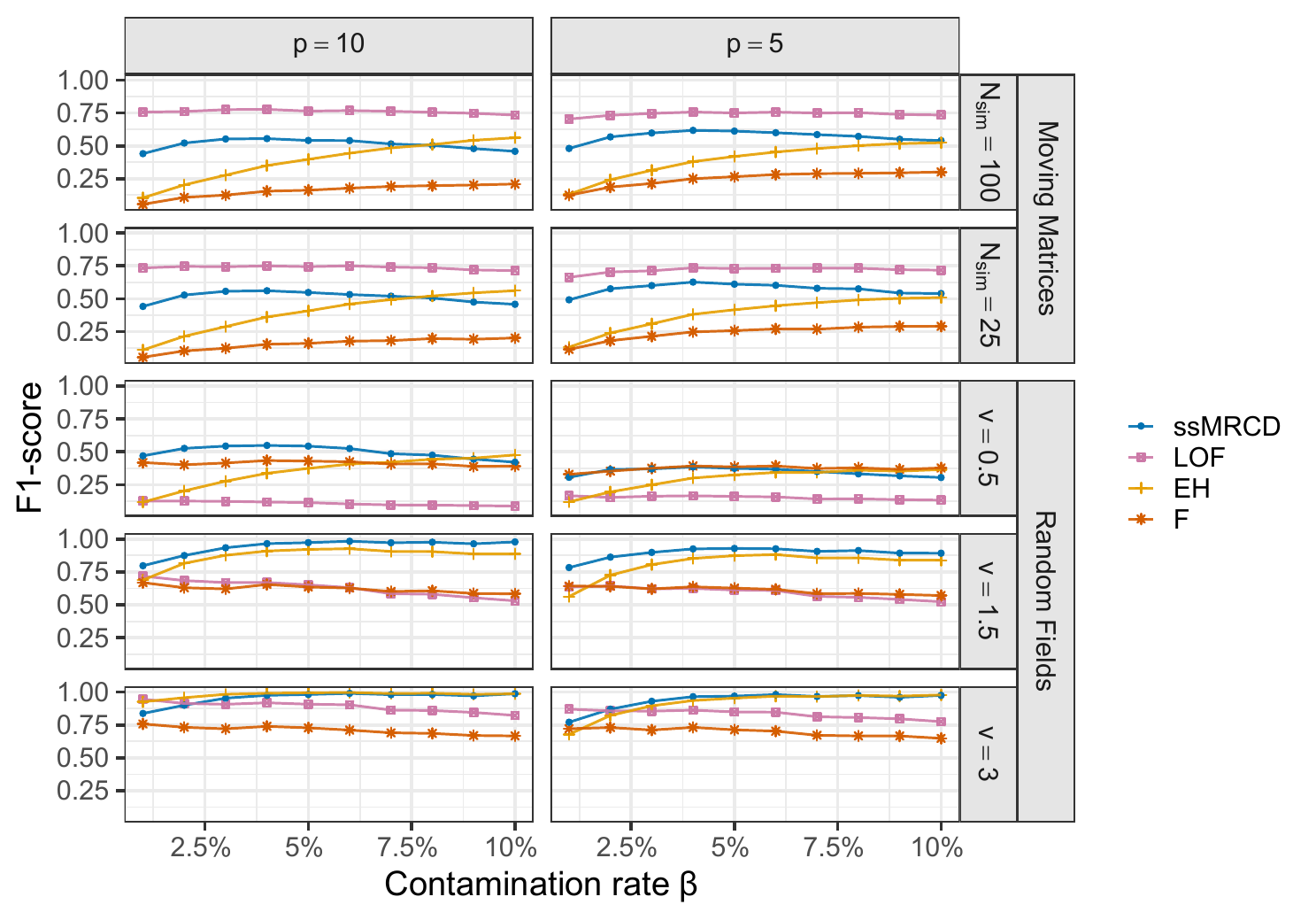}
    \caption{F1-Score for all four outlier detection methods with contamination achieved through completely random switching for different scenarios. Each point represents the mean of 100 repetitions.}
    \label{fig:Harris2}
\end{figure}

\begin{figure}
    \centering
    \includegraphics[width = 0.95\textwidth, trim = {0 0.6cm 0 0}, clip]{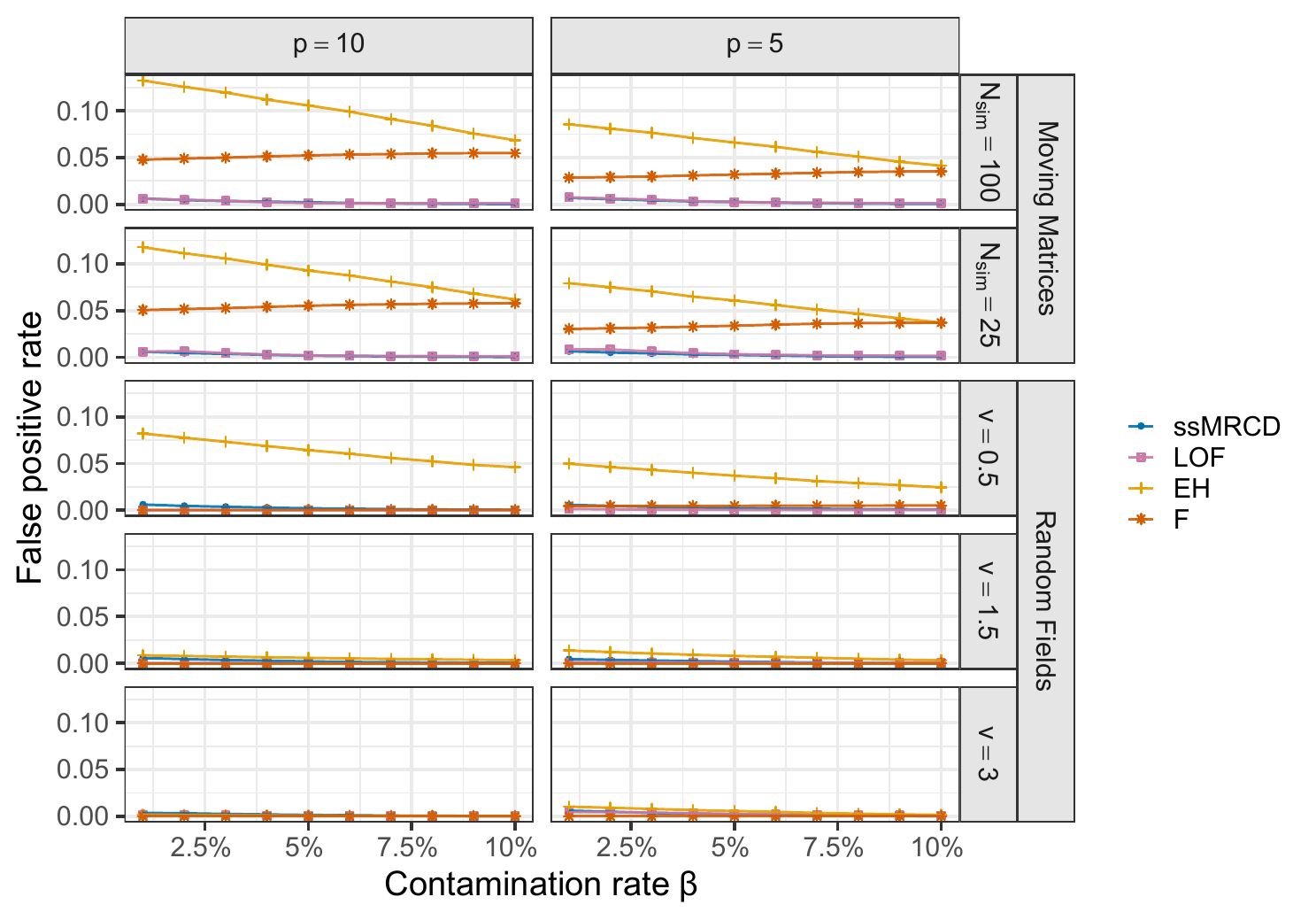}
    \includegraphics[width = 0.95\textwidth]{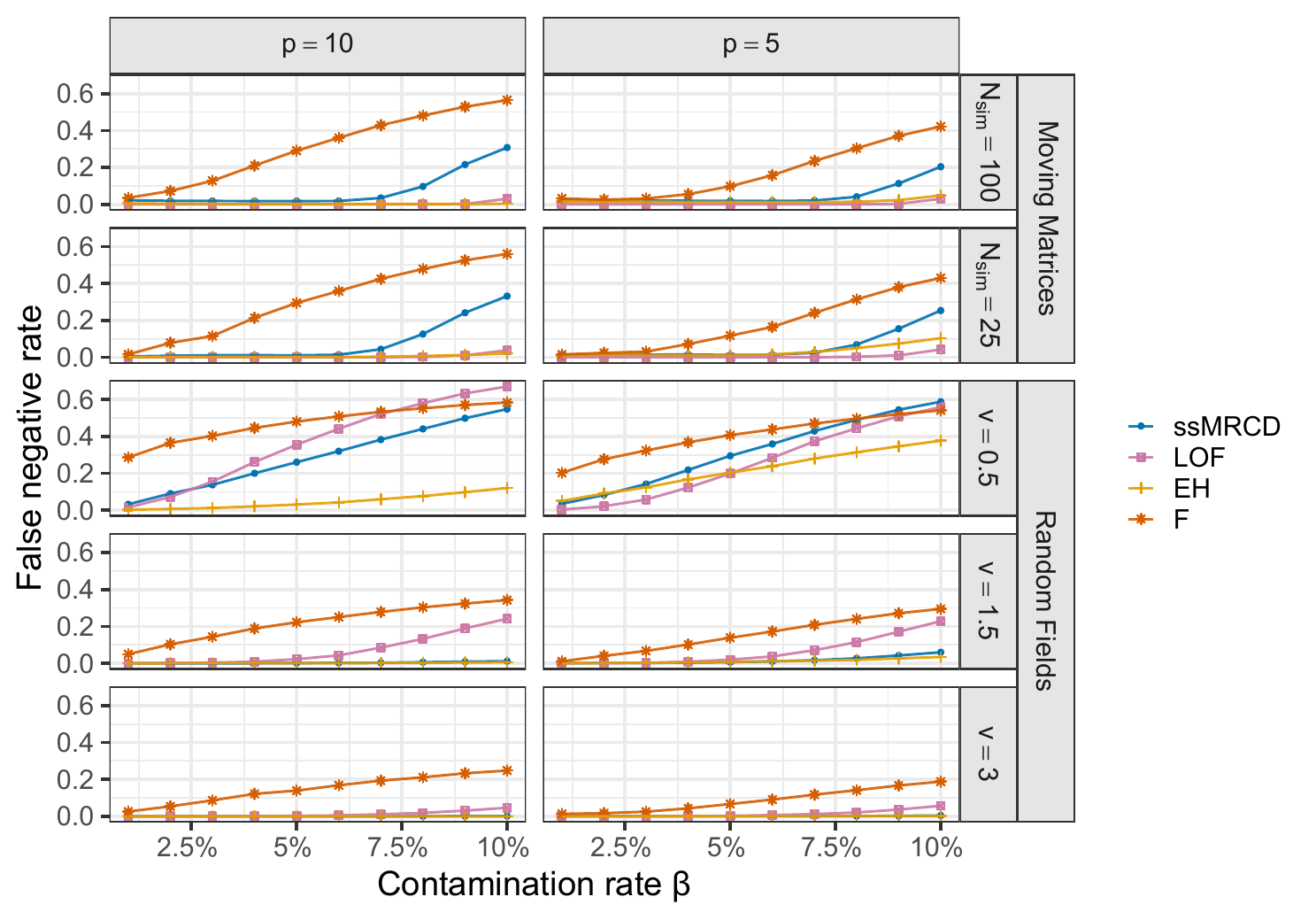}
    \caption{False positive and false negative rate for all four outlier detection methods with contamination achieved through the switching method of \cite{Ernst2016} for different scenarios. Each point represents the mean of 100 repetitions.}
    \label{fig:EH1}
\end{figure}

\begin{figure}
    \centering
    \includegraphics[width = 0.95\textwidth]{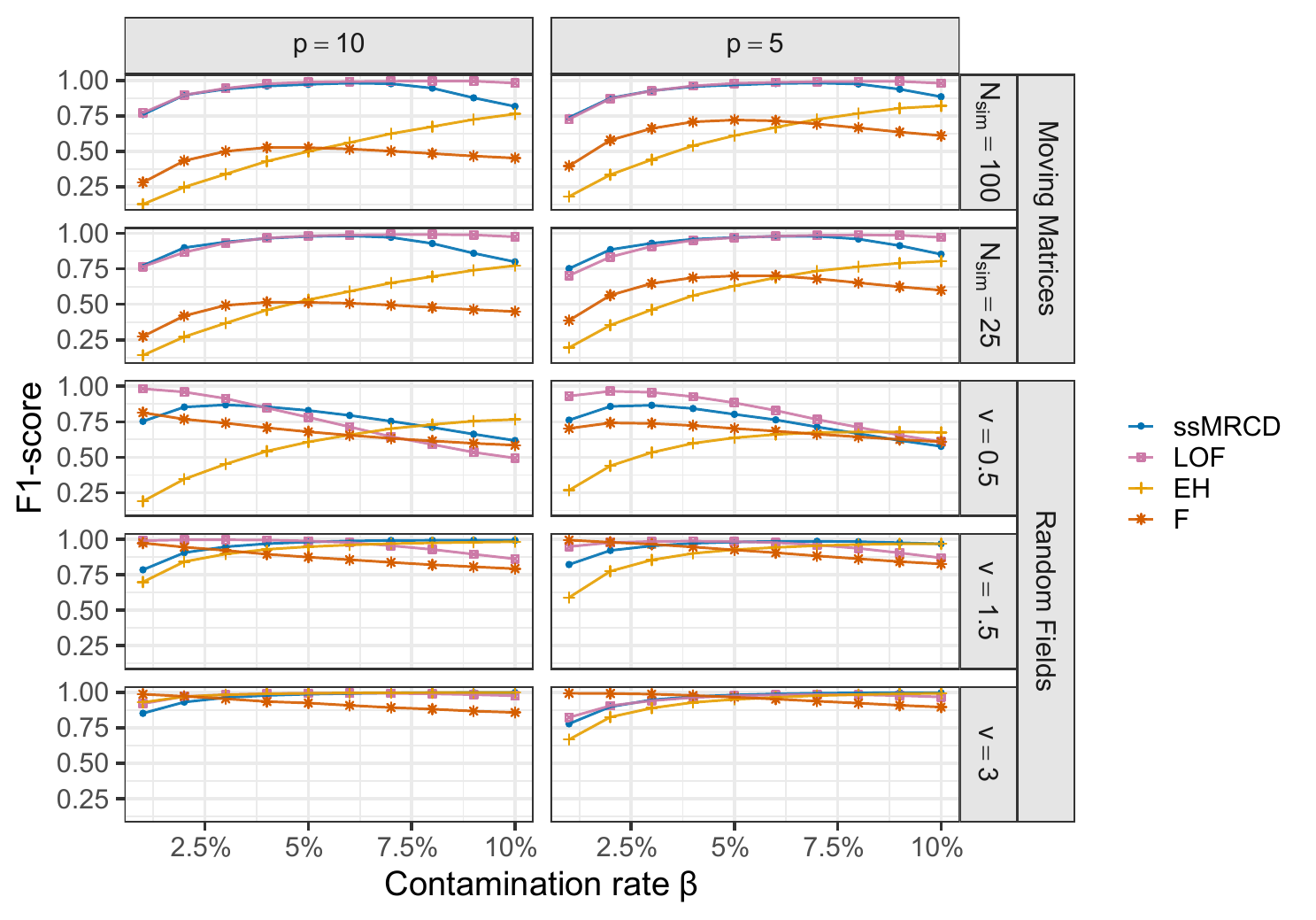}
    \caption{F1-Score for all four outlier detection methods with contamination achieved through the switching method of \cite{Ernst2016} for different scenarios. Each point represents the mean of 100 repetitions.}
    \label{fig:EH2}
\end{figure}

The results for both simulation setups and the completely random switching are shown in Figures~\ref{fig:Harris1} and \ref{fig:Harris2}. The results for the swapping method of \cite{Ernst2016} are shown in Figures~\ref{fig:EH1} and \ref{fig:EH2}.

Starting with the false positive rate (FPR), we see that the method of \cite{Ernst2016} has some issues with classifying too many normal observations as outliers in nearly all settings. This is likely due to the very local covariance estimation which might be too strict in general leading to a strong swamping effect, especially in settings where there is no strong spatial correlation of the observed values. Interestingly, the behaviour of the FPR for the method of \cite{Filzmoser2013} depends on the data simulation setup. For the moving matrix setup, the FPR is rather high, for random fields it is very low. The LOF and the ssMRCD-based outlier detection method have reliable low FPR for all setups. 

The outcome for the false negative rate (FNR) is quite different. While for \cite{Ernst2016} the FNR is in many settings below all other methods, this might just be due to the high FPR. The method of \cite{Filzmoser2013} has a very high FNR in most scenarios even in those with a high FPR. Regarding LOF, the simulation setup has a strong effect on its performance. For the moving matrix setup we see a rather good performance compared to the other methods, while for random fields the FNR can hardly keep up with the other methods except the method of \cite{Filzmoser2013}. The ssMRCD method is somewhere in between. Although the corresponding FNRs for the moving matrix setup with completely random swapping are not overwhelming, they are still in a reasonable range for the switching method of \cite{Ernst2016}. Moreover, the FNRs for the ssMRCD outlier detection technique in the other scenarios are compellingly low.

Comparing the F1-scores of the different methods, the best method to use in general depends on the scenario. While the three selected methods seem to have pitfalls in at least one simulation setup, the ssMRCD-based method is consistently showing reliable results and is mostly among the two best methods. Moreover, less extreme behavior occurs when it comes to the FNR and the FPR. Thus, the ssMRCD-based local outlier detection method could be the method of choice for standard outlier detection tasks.

\begin{figure}[htp]
    \centering
    \includegraphics[width = 0.95\textwidth]{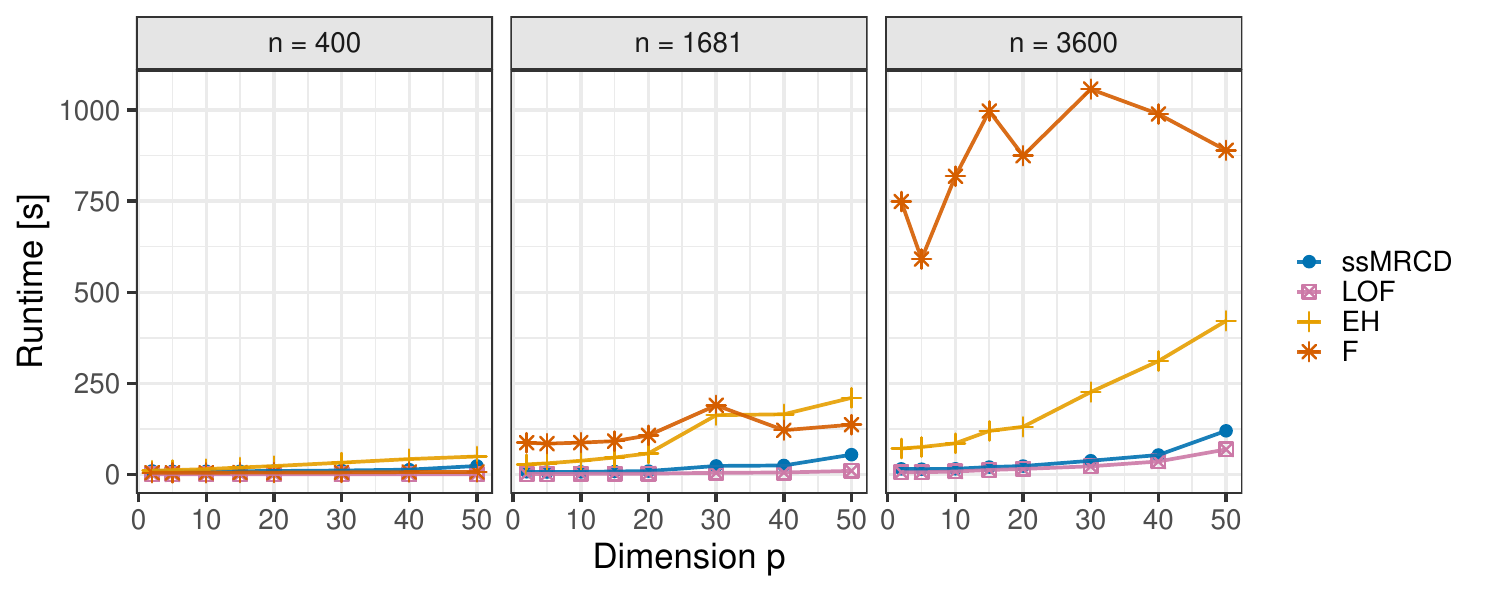}
    \caption{Comparison of average (mean) runtime for outlier detection algorithms using simulation setup 1 with varying number of observations $n$ and dimension $p$. The parameter settings from the previous outlier detection performance simulation study are used.}
    \label{fig:runtime_comparison}
\end{figure}

Moreover, with many dimensions and observations efficiency in computing becomes important. The results of a short simulation study regarding the runtime of the four outlier detection techniques with parameter settings according to the prior analysis are shown in Figure~\ref{fig:runtime_comparison} for the moving matrix scenario with $N_{sim} = 100$. All methods need more computation time for increasing dimension $p$, especially the methods based on covariance estimation and inversion (EH, ssMRCD, F). Also, the number of observations seems to have a big effect on runtime, especially for the method of \cite{Filzmoser2013}, possibly due to an inefficient implementation of finding the k-nearest neighbors. Although the local outlier factor (LOF) method is reliably fast, the ssMRCD seems to be comparably efficient, even though it involves a complex covariance estimation procedure. 

\section{Example}
\label{sec:examples}

\begin{figure}[htp]
    \centering
    \includegraphics[width = \textwidth]{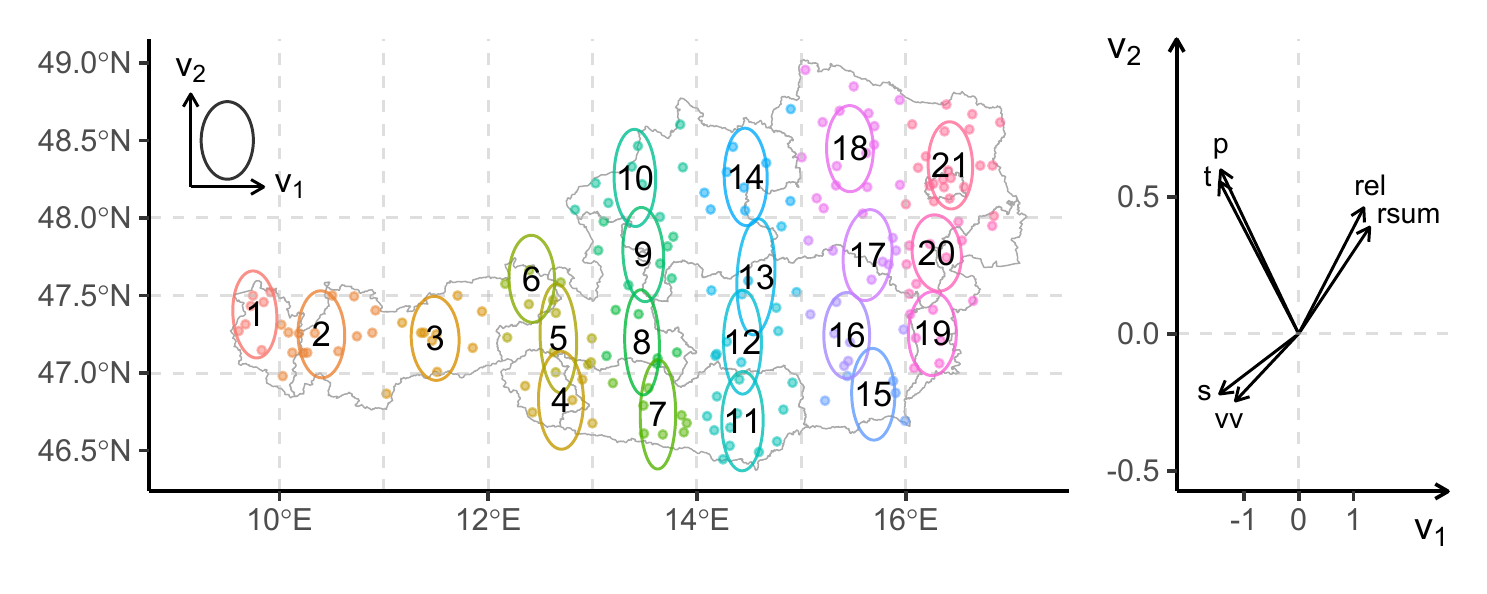}
    \caption{On the left hand side, the countours of Austria and its districts are shown together with the imposed grid structure for the ssMRCD neighborhoods. Here, singular points are assigned to a neighboring neighborhood. For each neighborhood $a_i$ the index $i$ is placed at the center, and the tolerance ellipses of corresponding correlation matrices are plotted along the first and second eigenvector coordinate of $\boldsymbol{T}$, which can be seen in the upper left corner as reference. The biplot of $\boldsymbol{T}$ is shown on the right hand side. The observations and ellipses are colorized according to their neighborhoods.}
    \label{fig:AUTcovariance}
\end{figure}

\begin{figure}[htp]
    \centering
    \includegraphics[width = \textwidth, trim = {0 1cm 0 0.5cm}, clip]{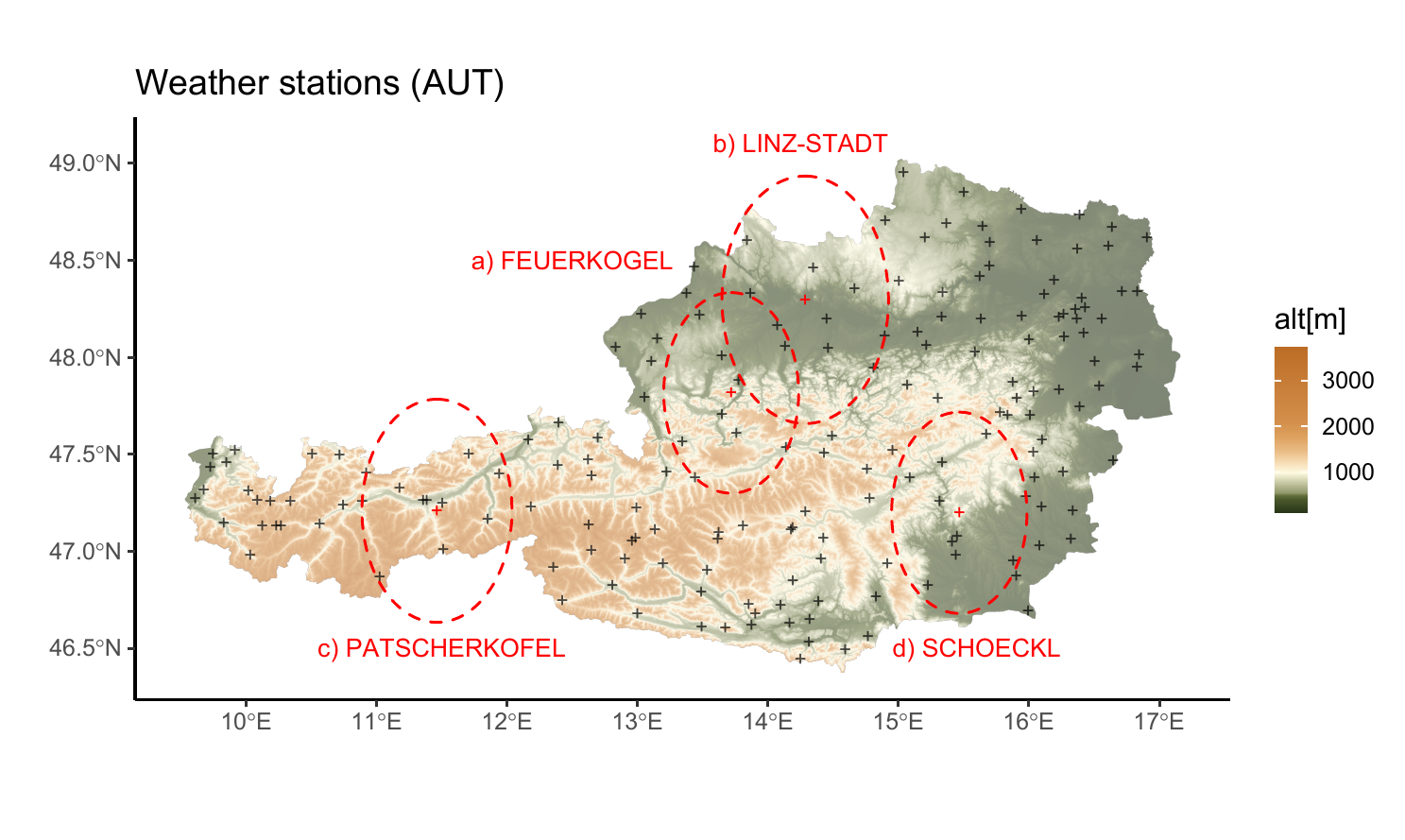}
    \caption{Altitude map of Austria with all weather stations and four selected outliers in red. Each red-dashed ellipse indicates the $k$ nearest neighbors with whom the outlier is compared.}
    \label{fig:AUTmap}
\end{figure}

In this section we consider a data set provided by \cite{ZAMG2022}. It  consists of monthly weather data for Austrian weather stations and is used to test and compare the different methods. The data set contains measurements of air pressure [hPa] (p), relative air humidity [\%] (rel), the monthly sum of sunshine duration [h] (s), wind velocity [m/s] (vv), air temperature in 2 meters above the ground [°C] (t), and the average daily sum of precipitation [mm] (rsum), averaged over all months in 2021, for $n = 183$ weather stations distributed all over Austria (see also Figure~\ref{fig:AUTcovariance} or \ref{fig:AUTmap}). The coordinates used for all methods are given in latitude and longitude.

We set $k = 10$ for all methods, i.e.~we want to compare one observation with its ten closest neighbors, independent of the methodology used. Although for the method of \cite{Ernst2016} it is possible to remove observations with comparably high levels of heterogeneity among the neighbors, we want to include all observations, thus setting $\beta = 1$. Even if there is increased heterogeneity among the neighbors, an observation might still be an interesting outlier clearly visible with the naked eye. Moreover, as mentioned in the prior section, the simulation results in \cite{Ernst2016} show the best performance for high $\beta$. For the methodology of \cite{Filzmoser2013} in accordance with the simulation setup we allow for one of the ten neighbors to be similar to the local outlier. The cutoff value is again set to $0.2$. We use the same cutoff value of $1.5$ for the LOF as in the simulations.

For the ssMRCD local outlier detection method we use a grid based neighborhood structure for the covariance estimation. Due to the Alpine landscape especially in the Western parts of Austria we aim at a rather local covariance structure, thus choosing a rather fine grid with $N = 21$ neighborhoods and $n_i \approx 8.7$ observations per neighborhood on average. Other possible options could be based on underlying structures, e.g. due to historical or political reasons, or on other classifying methods like clustering of the spatial coordinates. Furthermore, we use inverse-distance weights for the weighting matrix $\boldsymbol{W}$ between neighborhoods based on their center and select the default smoothing degree of $\lambda = 0.5$ to gain enough smoothing but still keep the locality of the fine grid structure. 

As an alternative to using the default value of $\lambda = 0.5$ we can set up a simulation procedure. Assuming that the real data is uncontaminated, we can swap observations similar to the simulation studies in Section~\ref{sec:Numerical_simulations} and define them as local outliers. We can apply the outlier detection technique with the ssMRCD and different choices of $\lambda$ (this can also be applied to other parameter settings of the ssMRCD), and then analyse the fraction of found outliers and the total number of outliers. Since only focusing on the known FNR for the found outliers leads to an increased false positive rate, it is sensible to also take the total number of found outliers into account. A good value of $\lambda$ is a trade-off between a low FNR and a comparatively low number of found outliers overall. Interestingly, this procedure endorses the choice of $\lambda = 0.5$ in this data set.

The resulting ssMRCD correlation matrices for each neighborhood can be seen in Figure \ref{fig:AUTcovariance}. The observations and the tolerance ellipses of the ssMRCD correlation matrices are colorized according to their neighborhoods. Since we have dimension $p=6$, we reduce the dimensionality to the first two eigenvectors $v_1$ and $v_2$ of the global MCD correlation matrix $\boldsymbol{T}$, which is displayed at the upper left corner, hoping to depict most of the relevant variance. Moreover, a biplot of $\boldsymbol{T}$ is added at the right hand side to link the correlation matrices to our weather data.

\begin{figure}[htp]
    \centering
    \includegraphics[width = 1.1\textwidth]{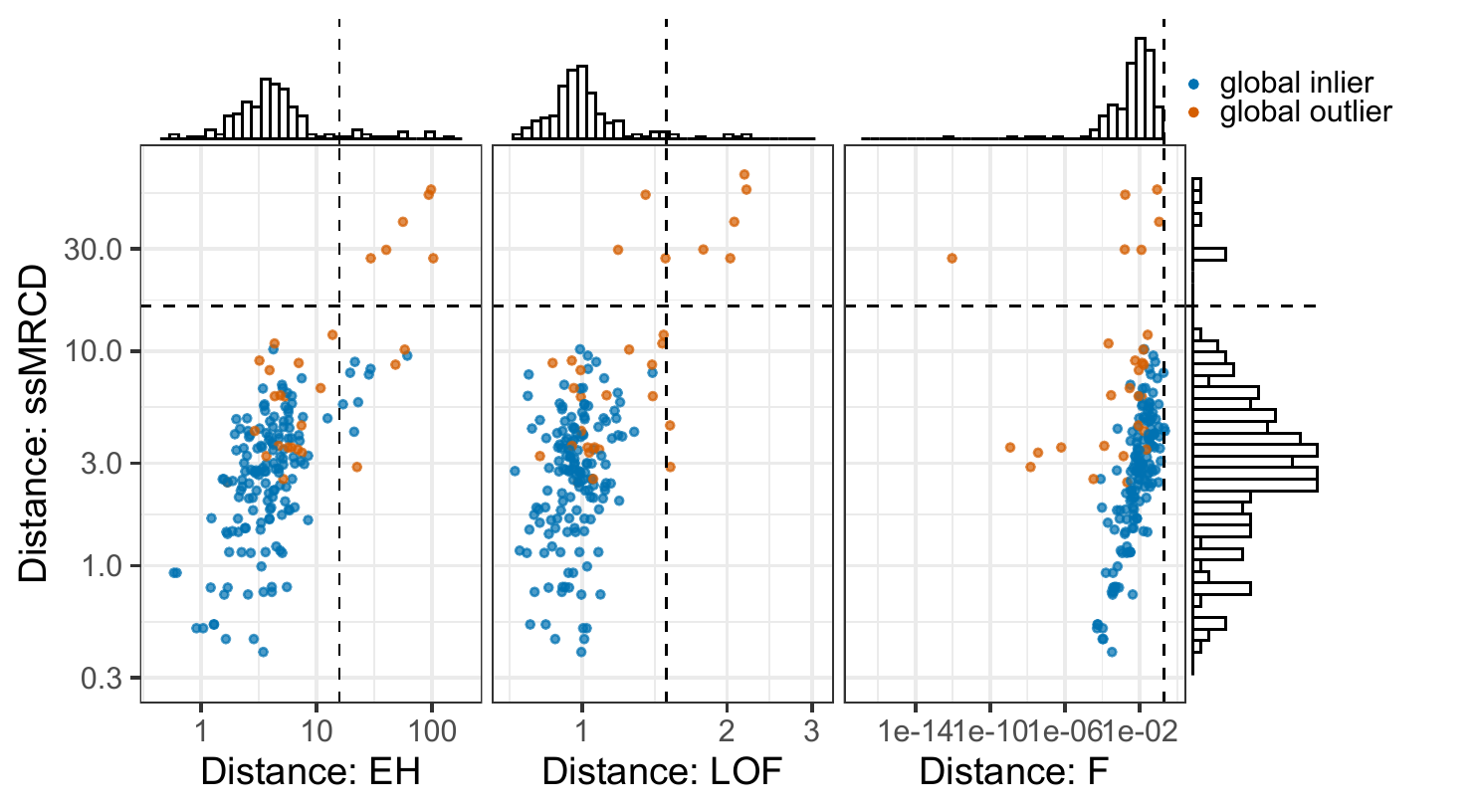}
    \caption{Distance-distance plots with the outlyingness scores of EH (next distance), of LOF (local outlier factor) and of F (isolation degree) against the next distance of ssMRCD. Observations colored in orange are global outliers based on the robust MD with the MCD as covariance estimator. At the margins the distribution of the different outlyingness scores are depicted by histograms.}
    \label{fig:AUTdistancedistance}
\end{figure}

\begin{figure}[htp]
    \centering
    \includegraphics[width = 1.1\textwidth]{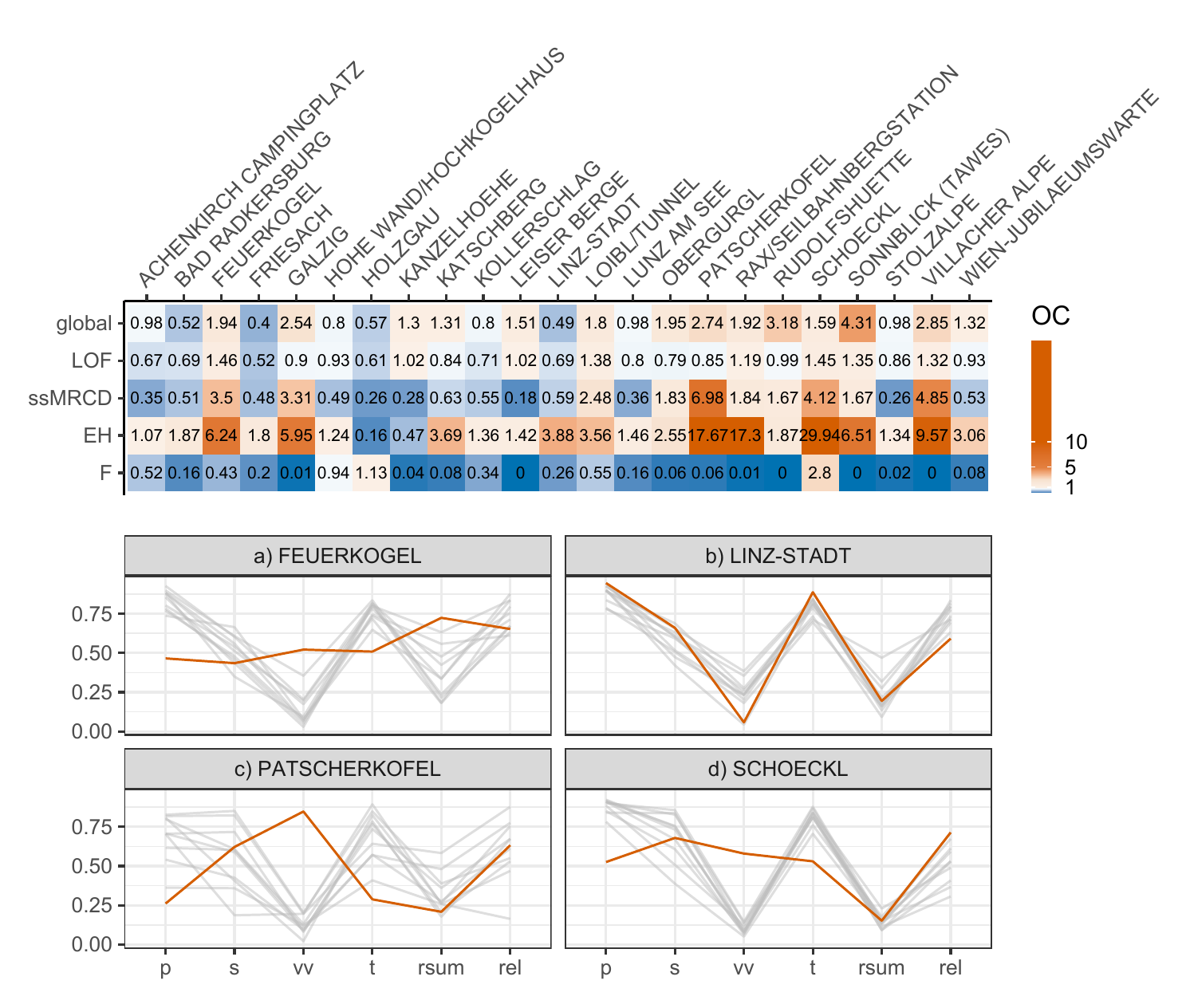}
    \caption{Upper part: Weather stations classified as outliers colorized according to their outlyingness score in relation to the cut-off value (OC) for all four methods and their global outlyingness. Lower part: in panels a)-d) four exemplary weather stations are selected to show differences on methodologies. Each variable is scaled to range $[0,1]$. The values of the selected outliers are colourized in orange, the corresponding $10$ nearest weather stations in grey.}
    \label{fig:AUTheatmap}
\end{figure}

Applying all four outlier detection methods leads to 24 observations in total classified as outliers. The most outliers (21) are classified by the method of \cite{Ernst2016}, the least (3) by \cite{Filzmoser2013}, which is consistent with the simulation results regarding FPR and FNR, especially for the random fields setup. The distances which are used for outlier detection for each method and observation are shown in Figure~\ref{fig:AUTdistancedistance}. For further comparison of the results, the upper part of Figure~\ref{fig:AUTheatmap} shows all 24 classified outliers with the corresponding ratio of distance value to cut-off value. Ratios above one are outliers. We can see that there are multiple weather stations that are classified as outliers only by the method of \cite{Ernst2016} which lends itself to a notion consistent with the simulation results that there are some false positives among these weather stations. One example for a false positive could be panel b) in in the lower part of Figure~\ref{fig:AUTheatmap}. The station Feuerkogel (panel a)) was not detected by the method of \cite{Filzmoser2013}, also consistent to the simulation results for the random fields setup and the generally high FNR. Interestingly, also LOF seems to have drawbacks and fails for example for the weather station Patscherkofel (panel c)), which was not detected as outlier. Nevertheless, the weather station Schoeckl (panel d)) was detected by all of them.

When looking at Figure~\ref{fig:AUTmap} we can find some explanation for the local outlyingness for two of the three local outlier stations and why panel b) might not be outlying. While the stations Schoeckl and Feuerkogel are rather exposed on higher altitudes than most of the surrounding k-nearest stations which can easily lead to different patterns regarding weather, the station Linz-Stadt (panel b)) is in a rather flat area similar to its neighbors. The station Patscherkofel is already deep in the Alpine area and is surrounded by other stations in valleys but also on mountains. Although from panel c) in Figure~\ref{fig:AUTheatmap} it is evident that Patscherkofel differs significantly in wind velocity, it is not clear why it differs so much also from stations with similar altitude and exposure.

\section{Conclusions}
In this paper we enhance the limited toolbox for multivariate local outlier detection by extending the approaches of \cite{Filzmoser2013} and \cite{Ernst2016}. The developed ssMRCD based on the MRCD \citep{Boudt2020} bridges the gap between fully local and fully global covariance matrices used in the pairwise MD by exchanging the extremely local covariance matrices used in \cite{Ernst2016} with spatially smooth estimates. 

We define the ssMRCD by means of a minimization problem and prove theoretical properties of the estimator, such as equivariance and breakdown point. A heuristic is provided for the stable convergence property of the proposed algorithm under reasonable spatial changes in underlying covariance matrices. Moreover, the methods of \cite{Filzmoser2013}, \cite{Ernst2016} and the ssMRCD outlier detection method are compared with the local outlier factor adapted for local outliers \citep{Schubert2012} regarding outlier detection performance and computational efficiency for simulated data and real world data from Austrian weather stations. 

While we support the conclusion of \cite{Ernst2016} that it is difficult to select the "best" method for outlier detection techniques, the ssMRCD-based outlier detection technique seems to be the only method providing reliable (but still improvable) results over all simulation scenarios. Additionally, it is able to compete with the other methods regarding runtime even though the computation is quite complex. However, for a thorough real data analysis it is still preferable to use different outlier detection methods and compare the results in order to exploit all possible advantages of the available methods. Comparing results of multiple methodologies provides more insight in the data and significant local outliers can be classified with more reliability overall.

The ssMRCD covariance structure can be exploited also beyond local outlier detection. All covariance based methods that are sensible to adapt to spatial data can be extended by using the ssMRCD instead, e.g. spatial principal component analysis. A special case for the application of the ssMRCD might also be spatial data with structural breaks that need to be considered in the analysis.
Finally, the presented ideas could also be transferred to a
time series context, where the spatial dependency is replaced by
the temporal dependency of multivariate time series, and the 
dependence structure could change over time. Such settings are usually quite challenging for outlier detection.


\section*{Acknowledgements} 

Co-funded by the European Union (SEMACRET, Grant Agreement no. 101057741) and UKRI (UK Research and Innovation).


 \bibliography{BIB_OutlierDetection}
 \bibliographystyle{apalike}

\end{document}